%% file: dmx_v4.tex
\documentclass[a4paper,11pt]{article}
\pdfoutput=1 

\usepackage{jheppub_ex}             
\usepackage[all]{hypcap}            
\usepackage{slashed}                
\usepackage[usenames,table]{xcolor} 
\usepackage{feyndiag}               
\usepackage{parskip}                
\usepackage{multirow}               
\usepackage{array}                  
\usepackage[section]{placeins}      
\usepackage{enumitem}               
\usepackage[utf8]{inputenc}         
\usepackage{bbm}                    
\usepackage{afterpage}              

\newcommand{\irrep}[2][0]{\ensuremath{\mathrm{\textbf{#2}}}}
\def\met{E_T^{\rm miss}}

\graphicspath{{./plots/}}
\title{Simplified Phenomenology for Colored Dark Sectors}
\author[a]{Sonia~El~Hedri,}
\author[a]{Anna~Kaminska,}
\author[a]{Maikel~de~Vries,}
\author[b,c]{Jose~Zurita}
\affiliation[a]{PRISMA Cluster of Excellence \& Mainz Institute for Theoretical Physics, Johannes Gutenberg University, 55099 Mainz, Germany}
\affiliation[b]{Institute for Nuclear Physics (IKP), Karlsruhe Institute of Technology, Hermann-von-Helmholtz-Platz 1, D-76344 Eggenstein-Leopoldshafen, Germany}
\affiliation[c]{Institute for Theoretical Particle Physics (TTP), Karlsruhe Institute of Technology, Engesserstra{\ss}e 7, D-76128 Karlsruhe, Germany} 
\emailAdd{elhed001@uni-mainz.de}
\emailAdd{akaminsk@uni-mainz.de}
\emailAdd{mdevrie@uni-mainz.de}
\emailAdd{jose.zurita@kit.edu}
\preprint{MITP/17-002, TTP17-006}
\arxivnumber{1703.00452}
\keywords{Phenomenological Models}

\abstract{\input{sections/abstract_v8}}

\begin{document}

\maketitle
\clearpage

\input{sections/introduction_v13}
\input{sections/model_v15}
\input{sections/relic_density_v15}
\input{sections/collider_v15}
\input{sections/conclusions_v8}
\input{sections/acknowledgements_v5}

\appendix
\input{sections/thermal_eq_v7}
\input{sections/bound_states_v7}

\bibliographystyle{JHEP}
\bibliography{dmx_v4}

\end{document}

%% file: sections/abstract_v8.tex
We perform a general study of the relic density and LHC constraints on simplified models where the dark matter coannihilates with a strongly interacting particle X. In these models, the dark matter depletion is driven by the self-annihilation of X to pairs of quarks and gluons through the strong interaction. The phenomenology of these scenarios therefore only depends on the dark matter mass and the mass splitting between dark matter and X as well as the quantum numbers of X. In this paper, we consider simplified models where X can be either a scalar, a fermion or a vector, as well as a color triplet, sextet or octet. We compute the dark matter relic density constraints taking into account Sommerfeld corrections and bound state formation. Furthermore, we examine the restrictions from thermal equilibrium, the lifetime of X and the current and future LHC bounds on X pair production. All constraints are comprehensively presented in the mass splitting versus dark matter mass plane. While the relic density constraints can lead to upper bounds on the dark matter mass ranging from $2$~TeV to more than $10$~TeV across our models, the prospective LHC bounds range from $800$ to $1500$~GeV. A full coverage of the strongly coannihilating dark matter parameter space would therefore require hadron colliders with significantly higher center-of-mass energies.

%% file: sections/introduction_v13.tex
\section{Introduction}
\label{sec:introduction}
A major enterprise for high-energy physics is to elucidate the nature of dark matter (DM). Although its existence is supported by a vast amount of experimental data informing on its long-range interactions (gravity), little is known on its particle properties, except that it is most likely neutral under electromagnetism and the strong force~\cite{Agashe:2014kda}. Since the current astrophysical observations can not be accommodated within the Standard Model~\cite{Bertone:2004pz}, DM is necessarily new physics (NP).

A bottom-up approach to dark matter model building uses simplified models~\cite{Abercrombie:2015wmb,Liew:2016oon,An:2013xka,DiFranzo:2013vra,Abdallah:2014hon,Buckley:2014fba,Harris:2014hga,Garny:2015wea,Abdallah:2015ter,Cirelli:2005uq}, that capture the phenomenology using a DM field and sometimes a mediator with a few NP couplings. The measurement of the relic abundance performed by the Planck satellite~\cite{Ade:2015xua} naturally sets a TeV scale mass for the DM candidate with weak scale interactions. For instance, in the well-studied MSSM case, pure Bino, Higgsino and Wino dark matter candidates require $m_\mathrm{DM} = 0.1, 1.1$ and $2.7$ TeV~\cite{Cirelli:2007xd}. Yet, it is important to stress that this minimal approach is mainly driven by simplicity. Stringent constraints coming from collider searches and direct detection experiments put these simplified models under siege~\cite{Beniwal:2015sdl,Malik:2014ggr,Alves:2013tqa,Alves:2015pea,Kahlhoefer:2015bea,Akerib:2013tjd,Aprile:2013doa,Cohen:2013ama}, inviting to explore new directions in dark matter model building.

Coannihilation of DM with neighbouring states is an ubiquitious feature in NP models that severely affects the relic density prediction~\cite{Griest:1990kh}. In that spirit, the Coannihilation Codex~\cite{Baker:2015qna} contains a systematic and complete classification of all simplified models featuring coannihilation, namely involving the process $\mathrm{DM} \, \mathrm{X} \to \mathrm{SM}_1 \, \mathrm{SM}_2$ at the renormalizable level. The Codex also features a generic study of all the collider signatures stemming from this setup. The inclusion of the $\mathrm{X}$ field renders the relic density a more complicated observable, now driven by several independent parameters and where many different processes contribute.  If $\mathrm{X}$ is charged under $SU(3)$, the collider phenomenology of these \emph{colored dark sectors}~\cite{Ellis:2015vaa,Liew:2016hqo,Ibarra:2015nca,deSimone:2014pda} will be dominated by the $p \, p \to \mathrm{X} \, \overline{\mathrm{X}}$ process and the relic density by $\mathrm{X} \, \overline{\mathrm{X}} \to \mathrm{SM}\, \mathrm{SM}$. The rate for both processes is purely determined by the strong gauge coupling, the X mass and its color representation. Hence there is a mild to negligible dependence on the NP couplings, which allows to set generic constraints on these scenarios. An accurate determination of the relic density also requires the proper inclusion of the Sommerfeld effect~\cite{Sommerfeld:1931aa,Cirelli:2007xd,deSimone:2014pda,Cassel:2009wt,Cassel:2009pu,Iengo:2009ni}, which was analyzed in detail in~\cite{ElHedri:2016onc} and the bound state formation~\cite{Ellis:2015vaa,Liew:2016hqo,Mitridate:2017izz}. In this paper we extend the existing results in the literature to also include the case of $\mathrm{X}$ being a vector and/or a color sextet. Moreover, we take into account constraints on the parameter space of the simplified models resulting from the requirement of thermal equilibrium in the early Universe and prompt decays of $\mathrm{X}$ at the LHC.

One of the main goals of this paper is to estimate how far beyond the LHC reach a dark matter candidate can be. This naturally sets the mechanism of coannihilation with a strongly interacting partner as the focus of our study. We consider simplified models where $\mathrm{X}$ is colored and the dark sector interacts with the SM via a higher dimensional operator suppressed by a scale $\Lambda \gtrsim 10$ TeV, effectively putting the integrated-out mediator beyond the LHC reach. Since $\mathrm{X}$ is charged under $SU(3)$, at least one of the SM particles it decays to has to be a quark or a gluon. Hence $\mathrm{X} \, \overline{\mathrm{X}}$ production via strong interactions always allows to test the colored dark sector in jet(s) plus transverse missing-energy collider searches. Considering other kinds of SM particles as $\mathrm{X}$ decay products would open up other search strategies including for example soft leptons, see~\cite{Giudice:2010wb,Rolbiecki:2012gn,Schwaller:2013baa}. Since we are interested in conservative LHC prospects, we consider both $\mathrm{SM}_{1,2}$ to manifest as jets. In this scenario the collider phenomenology is largely dominated by the multi-jet plus large missing transverse energy (MET) searches~\cite{ATLAS-CONF-2016-095,CMS-PAS-SUS-15-005,ATLAS-CONF-2016-078,CMS-PAS-SUS-16-014,CMS-PAS-SUS-16-016,Aaboud:2016zdn,Aaboud:2016tnv,Khachatryan:2016kdk}. The sensitivity to direct and indirect dark matter searches is very weak, as discussed in~\cite{Buschmann:2016hkc}. The relic density prediction is dominated by $\mathrm{X}$ annihilation, which opens up seemingly excluded parameter space in $\mathrm{DM}$ mass or allows for less fine-tuning in the $\mathrm{DM} - \mathrm{X}$ mass splitting than for electroweak coannihilation models. For simplicity we will assume DM to be a SM singlet, but our results do not depend on this choice. Since $\mathrm{X}$ needs to eventually decay into DM, requiring prompt X decays at the LHC naturally sets a lower bound on the effective coupling between the SM and the dark sector or on the $\mathrm{DM} - \mathrm{X}$ mass splitting. 

For each model, we also perform a detailed analysis of the collider phenomenology. We first recast the existing bounds on our scenario coming from monojet and multi-jet plus transverse missing-energy searches and present the parameter space allowed by the current data. Later, we extrapolate these bounds to study the reach of the HL-LHC with 3000 fb$^{-1}$ under two different scenarios for the systematic uncertainties, which are currently the bottleneck in many of these searches. We find that the LHC is currently probing masses between 300 GeV and 1 TeV, and the HL-LHC will further extend its reach to 700 GeV to 1.5 TeV. While this is a strong improvement, it is still not enough to fully cover the parameters space favored by thermal production of coannihilating dark matter, and hence we conclude that a higher center-of-mass energy collider would be necessary to probe the thermal region. We also allude to the capability of a future 100~TeV proton-proton collider~\cite{Golling:2016gvc,Contino:2016spe} to fully test these models.

The present paper is organized as follows. In section~\ref{sec:dm:partner} we discuss the models we consider. In section~\ref{sec:relic:density} we present the calculation of the relic density with a detailed treatment of the Sommerfeld effect (see~\cite{ElHedri:2016onc} for further details) and bound state formation. In section~\ref{sec:collider:pheno} we present the results of our collider analysis. We conclude in section~\ref{sec:conclusions}.

%% file: sections/model_v15.tex
\section{Simplified dark matter coannihilation}
\label{sec:dm:partner}
We consider a set of minimal models where the DM field is a pure Standard Model singlet and has no self-annihilation channels. Such scenarios can be made viable under the thermal hypothesis by introducing a coannihilation partner X, in thermal and chemical equilibrium with the dark matter particle. Provided that X and DM are very close in mass, the dark sector particles can then deplete via either the X\,DM~$\rightarrow$~SM$_1$\,SM$_2$ channel or the X\,$\overline{\mathrm{X}}$~$\rightarrow$~SM\,SM self-annihilation channel. In this study, we focus on the particular case of X being a colored particle. In this scenario, X will annihilate to quark and gluon pairs via QCD interactions. Since these annihilation processes involve strong couplings, they are likely to be the main drivers of the dark matter depletion even when other processes, notably electroweak processes, are present. 

The minimal Lagrangians for X being either a scalar $S$, a fermion $\psi$ or a vector $V_\mu$ are of the form
\begin{equation} \label{eq:lagrangians:x}
	\begin{aligned}
		\mathcal{L}_S & = \left[ D_{\mu,ij}^{\vphantom{\mu}} S_j\right]^\dagger \left[D_{ij}^\mu S_j \right] - m_S^2 \, S_i^\dagger S_i \\
		\mathcal{L}_\psi & = \bar{\psi}_i i \slashed{D}_{ij} \psi_j - m_\psi \bar{\psi}_i \psi_i \\
		\mathcal{L}_V & = -\frac{1}{2} {V_{\mu\nu,i}}^\dagger V^{\mu\nu}_i - i g_s {V_i^\mu}^\dagger (T^a_{\irrep{R}})_{ij} V_j^\nu G_{\mu\nu}^a + m_V^2 V_{\mu,i}^\dagger V^\mu_i ,
	\end{aligned}
\end{equation}
where $i, j$ are color indices and the $T^a_{\irrep{R}}$ matrices are the generators for the color representation $\irrep{R}$ of X. Note that these Lagrangians are for a complex scalar, a Dirac fermion and a complex vector; to obtain the Lagrangians for real scalars, Majorana fermions and real vectors each of the individual terms need to be multiplied by a factor of one half. The covariant derivatives and field strength are given by
\begin{equation}
	\begin{aligned}
		V^{\mu\nu}_i & = D^\mu_{ij} V^\nu_j - D^\nu_{ij} V^\mu_j \\
		D_{\mu, ij} & = \partial_\mu \delta_{ij} - i g_s G_\mu^a (T^a_{\irrep{R}})_{ij} .
	\end{aligned}
\end{equation}
Here we choose to ignore additional pieces such as the anomalous terms introduced in~\cite{Blumlein:1996qp,Hewett:1993ks}. We also remain agnostic about the mass generation mechanism for vector fields, which can lead to issues with perturbative unitarity, as discussed in section~\ref{sec:vector:unitarity}. Neglecting self interactions for dark matter~\cite{Spergel:1999mh,Wandelt:2000ad,Kaplinghat:2015aga}, the Lagrangians for the DM fields are of the form
\begin{equation} \label{eq:lagrangians:dm}
	\begin{aligned}
		\mathcal{L}_{S_\mathrm{DM}} & = \partial_\mu S^\dagger_\mathrm{DM} \partial^\mu S_\mathrm{DM} - m_\mathrm{DM}^2 S^\dagger_\mathrm{DM} S_\mathrm{DM} \\
		\mathcal{L}_{\psi_\mathrm{DM}} & = \bar{\psi}_\mathrm{DM} i \slashed{\partial} \psi_\mathrm{DM} - m_\mathrm{DM} \bar{\psi}_\mathrm{DM} \psi_\mathrm{DM} \\
		\mathcal{L}_{V_{\mu\mathrm{DM}}} & = -\frac{1}{2} {V_{\mu\nu,\mathrm{DM}}}^\dagger V^{\mu\nu}_\mathrm{DM} + m_\mathrm{DM}^2 V^\dagger_{\mu, \mathrm{DM}} V_\mathrm{DM}^\mu .
	\end{aligned}
\end{equation}
Again as for the Lagrangians in equation~\eqref{eq:lagrangians:x} a factor of one half needs to be inserted if the dark matter field is self-conjugate. In order to build a viable theory of thermal dark matter, we need to enforce chemical and thermal equilibrium between X and DM, and need to introduce a decay channel for X. All these requirements can be fulfilled by demanding the existence of a single effective operator $\mathcal{L}_{\mathrm{DM} + \mathrm{X}} \propto$ X\,DM\,SM$_1$\,SM$_2$ for each model. The structure of these operators for the different models are further discussed in section~\ref{sec:dm:partner:eft}. The final Lagrangian will therefore be of the form
\begin{equation}
	\mathcal{L} = \mathcal{L}_\mathrm{X} + \mathcal{L}_\mathrm{DM} + \mathcal{L}_{\mathrm{DM} + \mathrm{X}} .
\end{equation}
We assume that the dark matter field is protected by a global discrete symmetry similar to the $\mathbb{Z}_2$ parity. The coannihilation partner will have the same parity as dark matter under this symmetry and the pair will together form the dark sector.

The final set of models can be described by three parameters, namely $m_\mathrm{DM}$, $m_\mathrm{X}$ and the suppression scale $\Lambda$ of the $\mathcal{L}_{\mathrm{DM} + \mathrm{X}}$ operator together with three discrete choices for the spin and color of X and the spin of the dark matter. In what follows, we consider DM and X to be either real/complex scalars, Dirac/Majorana fermions or real/complex vectors and study color representations of X ranging over $\irrep{3}$, $\irrep{6}$, $\irrep{8}$. Note that higher representations of $SU(3)$ --- $\irrep{10}$, $\irrep{15}$ and $\irrep{27}$ --- are possible as well, however, it is difficult to realize these models in complete BSM theories. We assume that the $\mathcal{L}_{\mathrm{DM} + \mathrm{X}}$ interaction is suppressed, with $\Lambda = 10$~TeV so that the integrated-out fields lie beyond the reach of the LHC.

The total set of models comprises 72 discrete choices for the spins of the dark matter and its coannihilation partner, and the color representation of X. In the rest of this paper we investigate a representative subset of scenarios that highlight the dependence of the dark matter relic density and collider phenomenology on the quantum numbers of X. We study the following models
\begin{equation} \label{eq:model:sets}
	\begin{aligned}
		\mathrm{DM}_\mathrm{F} + \mathrm{X}_\mathrm{F3} & \qquad \qquad \mathrm{DM}_\mathrm{F} + \mathrm{X}_\mathrm{F6} && \qquad \qquad \mathrm{DM}_\mathrm{F} + \mathrm{X}_\mathrm{F8} \\
		\mathrm{DM}_\mathrm{S} + \mathrm{X}_\mathrm{C3} & \qquad \qquad \mathrm{DM}_\mathrm{S} + \mathrm{X}_\mathrm{F3} && \qquad \qquad \mathrm{DM}_\mathrm{S} + \mathrm{X}_\mathrm{W3} \, ,
	\end{aligned}
\end{equation}
where the subscripts denote the spin as S (real scalar), C (complex scalar), F (Dirac fermion), W (complex vector) and the color representation of X. The first three models in this list explore the dependence of the phenomenological constraints on the color representation of X, whereas the second three models illustrate the effect of changing the spin of the coannihilation partner. Note that by including models with a color sextet and/or massive vector bosons, this works expands the scope of the previous studies~\cite{Ellis:2015vaa,deSimone:2014pda,Liew:2016hqo} that were focusing on more traditional ``squark'' and ``gluino'' models.

Attached to this paper we ship a \texttt{FeynRules v2.3.24}~\cite{Christensen:2008py, Alloul:2013bka} package that contains all 72 models~\cite{deVries:2017xyz}. A \texttt{Mathematica} notebook has been added as well to extract each specific model in both \texttt{UFO}~\cite{Degrande:2011ua} and \texttt{CalcHEP v3.6.25}~\cite{Belyaev:2012qa} format. These can than be interfaced with \texttt{micrOMEGAs v4.3.2}~\cite{Belanger:2014vza} to calculate the relic abundance as well as with \texttt{MadGraph5 v2.5.2}~\cite{Alwall:2011uj, Alwall:2014hca} for the collider studies. These model files can be used in conjunction with the Sommerfeld corrections package~\cite{ElHedri:2016pac} we shipped with~\cite{ElHedri:2016onc}.

\subsection{Dark vectors and unitarity}
\label{sec:vector:unitarity}
In the Lagrangian shown in equation~\eqref{eq:lagrangians:x}, we introduced a St\"uckelberg mass for X~$=V_\mu$. Scenarios with a vector X will therefore lead to unitarity violation for the $\mathrm{X} \, \overline{\mathrm{X}} \rightarrow \mathrm{X} \, \overline{\mathrm{X}}$ and $\mathrm{X} \, \overline{\mathrm{X}} \rightarrow q \, \bar{q}$, $g \, g$ amplitudes at high center-of-mass energies. As for the Higgs mechanism in the Standard Model, unitarity can only be restored by introducing new particles that will be responsible for generating the mass of X.

We compute the maximal energy scale at which these new particles should appear by considering $\mathrm{X} \, \overline{\mathrm{X}} \rightarrow \mathrm{X} \, \overline{\mathrm{X}}$ scattering at high center-of-mass energy $s$. In this regime, the dominant contributions to the amplitude come from the longitudinal degrees of freedom and we can write
\begin{equation}
	\mathcal{A} \approx -\frac{\pi i \alpha_s}{m_\mathrm{X}^4} \left[ T^a_{ij} T^a_{kl} s(t - u) + T^a_{ik} T^a_{jl} t (s - u) + T^a_{il} T^a_{jk} u (t - s) \right] ,
\end{equation}
where $T^a$ is the color generator for the color representation of X and $i,j,k,l$ are the color indices of the initial and final state particles. This amplitude can be decomposed into partial waves $\mathcal{T}^J$ of the form
\begin{equation}
	\mathcal{T}^J = \frac{1}{32\pi}\sqrt{1 - \frac{4 m_\mathrm{X}^2}{s}} \int_{-1}^1 \mathcal{A}(\cos\theta) P_J(\cos \theta) \,\mathrm{d} \cos \theta ,
\end{equation}
where $\theta$ is the scattering angle and the $P_J(\cos\theta)$ are the Legendre polynomials. In the large $s$ regime, the zero-th partial wave can be approximated by
\begin{equation} \label{eq:unitarity}
	\begin{aligned}
		\mathcal{T}^0_{(ij)(kl)} &\approx \frac{1}{16 \pi} \int_{-1}^1 \mathcal{A}(\cos \theta) \,\mathrm{d} \cos \theta \approx \frac{2i \alpha_S s^2}{48 m_\mathrm{X}^4} \left[ T^a_{ik} T^a_{jl} - T^a_{il} T^a_{jk} \right] .
	\end{aligned}
\end{equation}
Equation~\eqref{eq:unitarity} allows to construct the $\mathcal{T}^0_{(ij)(kl)}$ matrix formed by all the possible pairs of color indices $(i,j)$. In order for a theory to be unitary, the eigenvalues of this matrix need to verify $|\lambda_i| < 1/2$. For our process, this constraint leads to an upper bound on the center-of-mass energy of the interaction that depends only weakly on the color representation of $\mathrm{X}$. This bound can in turn be translated into the following upper limit on the masses of the new particles needed to restore unitarity 
\begin{equation}
	M_\mathrm{NP}\sim \frac{\sqrt{s}}{2} \lesssim 2 m_\mathrm{X}. 
\end{equation}

Unitarity constraints therefore imply the existence of new particles at the same scale as X. Notably, in order to cancel the $\mathrm{X} \, \overline{\mathrm{X}} \rightarrow q \, \bar{q}$ divergences, it is necessary to introduce a fermionic quark partner $Q$ that couples to X and a Standard Model quark as well as a color-neutral gauge boson that couples to both $\mathrm{X} \, \overline{\mathrm{X}}$ and $q \, \bar{q}$. The quark partner can be either neutral or colored depending on the color of X. In order to account for the gauge boson masses, a complete theory should also involve a new Higgs-like multiplet with at least one color-neutral component $\phi$ that gets a vev and couples to two X bosons. An example of such a model, where the massive vector bosons arise from the breaking of $SU(4)$ to $SU(3)_c$, has been proposed in~\cite{Fornal:2015one}. Aside from the specific scenarios where either $Q$ is the dark matter or some of the new particles are very close in mass to the DM, introducing these extra particles will not lead to new (co)annihilation processes or additional decay modes for X. We should therefore expect the collider bounds on X to remain similar to the ones derived for a minimal model with only DM and X. Restoring unitarity will in fact lead to tighter constraints on most of the models involving a vector X since the LHC bounds on the masses of the additional particles might supplant the ones on the coannihilation partner.

Although merely restoring unitarity should not qualitatively change the relic density and collider studies presented in this work, requiring the new vector bosons to also be gauge bosons severely restricts their interactions with other particles. In fact generating an X\,DM\,SM$_1$\,SM$_2$ effective operator in models with massive dark vector bosons requires introducing a large number of new particles, which results in an elaborate model-dependent coupling structure. In the rest of this work, we will therefore not make any assumption about the structure of the X\,DM\,SM$_1$\,SM$_2$ operator for models with a vector X. We will still discuss, however, possible effective operators for models with scalar and fermion X in the next section.

\subsection{Effective operators}
\label{sec:dm:partner:eft}
In order to ensure chemical equilibrium between dark matter and X and to make X decay before BBN, we introduce a DM\,X\,SM$_1$\,SM$_2$ interaction term where SM$_1$ and SM$_2$ can be quarks or gluons depending on the quantum numbers of X and DM. The corresponding operator would be generated by either tree-level or loop interactions involving new physics at a scale $C \cdot g_\mathrm{NP}^m/\Lambda^n$, where $\Lambda$ is the suppression scale (in GeV), $g_\mathrm{NP}$ is a new coupling constant pertaining to the integrated out field, and $C$ is a numerical prefactor. In this work, we absorb the dependence in the coupling as well as the prefactors that are not loop factors into $\Lambda$.  The specific value of $\Lambda$ does not affect the phenomenology of the model as long as this scale is large enough  to ensure that the integrated out particles are outside the reach of the LHC and do not affect the dark matter depletion rate.\footnote{Additional particles that affect the decay of X and are within the reach of the LHC are still allowed as long as their couplings to the Standard Model, DM and X are suppressed.} We therefore set $\Lambda = 10$~TeV throughout this study. Note that choosing a higher value for this scale would increase the lifetime of X as well as slow down the DM~$\leftrightarrow$~X exchange process, which would further constrain the parameter space of our models. Indeed, such a higher scale would be needed when considering a future hadron collider with a larger center-of- mass energy. 

Since the coupling structure of the DM\,X\,SM$_1$\,SM$_2$ effective operator will not affect the relic density and collider constraints associated to the different models, we consider only the interaction terms with the lowest possible dimensionality. For the models under consideration, we choose the following operators, which have also been implemented in the \texttt{Feynrules} package~\cite{deVries:2017xyz}.\footnote{Although other coupling structures are allowed, we leave the construction of the corresponding interaction terms to the reader. We stress again that the choice of operator does not affect the dark matter annihilation rate or the collider phenomenology, but only the bounds associated with the lifetime of X (see section~\ref{sec:dm:partner:life}) and thermal equilibrium (see section~\ref{sec:thermal:eq}).}
\begin{equation} \label{eq:lagrangians:eft}
	\begin{aligned}
		\mathcal{L}_{\mathrm{DM}_\mathrm{F} + \mathrm{X}_\mathrm{F3}} & = \frac{1}{\Lambda^2} \epsilon_{kij} \left( \bar{\psi}_k \psi_{\mathrm{DM}} \right) \left( \bar{d}_{R,i} u^C_{R,j} \right) + \mathrm{h.c.} \\
		\mathcal{L}_{\mathrm{DM}_\mathrm{F} + \mathrm{X}_\mathrm{F6}} & = \frac{1}{\Lambda^2} K_{6,ij}^u \left( \bar{\psi}_{\mathrm{DM}} \psi^u \right) \left( \bar{u}_{R,i} u^C_{R,j} \right) + \mathrm{h.c.} \\
		\mathcal{L}_{\mathrm{DM}_\mathrm{F} + \mathrm{X}_\mathrm{F8}} & = \frac{1}{\Lambda^2} T_{ij}^a \left( \bar{\psi}_{\mathrm{DM}} \gamma_\mu \psi^a\right) \left(\bar{u}_{R,i} \gamma^\mu u_{R,j} \right) + \mathrm{h.c.} \\
		\mathcal{L}_{\mathrm{DM}_\mathrm{S} + \mathrm{X}_\mathrm{C3}} & = \frac{1}{\Lambda} \epsilon_{kij} \left( S_{\mathrm{DM}} S_k \right) \left( \bar{d}_{R,i} u_{R,j}^C \right) + \mathrm{h.c.} \\
		\mathcal{L}_{\mathrm{DM}_\mathrm{S} + \mathrm{X}_\mathrm{F3}} & = \frac{1}{16 \pi^2 \Lambda^2} T^a_{ij} S_{\mathrm{DM}} \left(\bar{d}_{R,i} \sigma^{\mu\nu} \psi_j \right) G^a_{\mu\nu} + \mathrm{h.c.} 
	\end{aligned}
\end{equation}
As mentioned in section~\ref{sec:vector:unitarity} we do not introduce any effective operator for the DM$_\mathrm{S}$ + X$_\mathrm{W3}$ model. Due to the unitarity requirement as well as the stringent restrictions on couplings involving gauge bosons, this model should involve a large number of new particles well below the scale $\Lambda$. Constructing a valid EFT for these scenarios is therefore not possible. In what follows, we assume that the new particles needed to complete the model will ensure that the lifetime and thermal equilibrium constraints discussed in sections~\ref{sec:dm:partner:life} and~\ref{sec:thermal:eq} are satisfied.

\subsection{Lifetime of X}
\label{sec:dm:partner:life}
In order for the models studied in this work to be viable, the colored coannihilation partner X needs to decay. In section~\ref{sec:dm:partner:eft}, we addressed this requirement by introducing an X\,DM\,SM$_1$\,SM$_2$ effective operator with its suppression scale $\Lambda$ fixed at $10$~TeV. Each of the interaction terms detailed in equation~\eqref{eq:lagrangians:eft} could a priori lead to a valid theory with X decaying before BBN. This requirement is however insufficient in sight of the current collider bounds. Long-lived coannihilation partners produced at colliders would form R-hadrons~\cite{Fayet:1977yc,Fayet:1978qc,Farrar:1978xj} that are constrained up to a few TeV by the current LHC searches~\cite{Aad:2013gva,Aad:2015rba,Aaboud:2016dgf,Aaboud:2016uth,Khachatryan:2015jha,Khachatryan:2016sfv}. In what follows, we therefore require the X decays to be prompt. Since the $\Lambda$ suppression scale is fixed in our study, this new requirement places constraints on the dark matter mass $m_\mathrm{DM}$ and the relative mass splitting $\Delta$ between DM and X. 

The probability for a particle with mass $m_\mathrm{X}$, momentum $p$, and decay width $\Gamma$ to travel a distance $d$ is given by an exponential distribution:
\begin{equation}
	P(d | p) = e^{-d/d_0(p)}\quad \text{with}\quad d_0(p) = \frac{\hbar c}{\Gamma}\frac{p}{m_\mathrm{X}} .
\end{equation}
The probability density for the particle to travel a distance $d$ can then be written as
\begin{equation} \label{eq:distance:proba}
	\mathcal{P}_d(d) = \int_0^\infty \frac{P(d|p)}{d_0(p_T, \eta, \phi)} \mathcal{P}_{p_T, \eta, \phi}(p_T, \eta, \phi) p_T \,\mathrm{d} p_T \,\mathrm{d} \eta \, \mathrm{d} \phi ,
\end{equation}
where $\mathcal{P}_{p_T,\eta,\phi}(p_T, \eta, \phi)$ is the probability density associated with the four-momentum of the particle.
In this study, we consider a particle to be long-lived when it is able to get out of the beam pipe. This requirement implies that the transverse distance $d_T$ traveled by the particle is larger than the beam pipe radius, which translates into
\begin{equation}
	d_T = d\sin\theta \ge d_\mathrm{beam} \sim 2.5~\mathrm{cm} .
\end{equation}
Here we introduced $\theta$, which is the angle between the particle track and the beam axis. Injecting this requirement into equation~\eqref{eq:distance:proba}, the probability for a particle to be long-lived is then
\begin{equation} \label{eq:distance:longlived}
	\begin{aligned}
		P(d_T > d_\mathrm{beam}) & = \int_{\frac{d_\mathrm{beam}}{\sin\theta}}^\infty \mathcal{P}_d(d)\,\mathrm{d} d \\
		& = \int_0^\infty \exp\left(-\frac{d_\mathrm{beam}}{d_0(p)\sin\theta}\right) \mathcal{P}_{p_T,\eta,\phi}(p_T, \eta, \phi) p_T \mathrm{d} p_T \mathrm{d} \eta \mathrm{d} \phi \\
		& = \int_0^\infty \exp\left(-\frac{d_\mathrm{beam}}{d_0^T(p_T)}\right) \mathcal{P}_{p_T} (p_T) p_T \mathrm{d} p_T ,
	\end{aligned}
\end{equation}
where we introduce the characteristic transverse distance
\begin{equation}
	d_0^T(p_T) = d_0(p) \sin \theta = \frac{\hbar c}{\Gamma} \frac{p_T}{m_\mathrm{X}} ,
\end{equation}
and the probability density for a particle to have a transverse momentum $p_T$
\begin{equation} \label{eq:proba:pt}
	\mathcal{P}_{p_T} = \int \mathcal{P}_{p_T, \eta, \phi}(p_T, \eta, \phi) \mathrm{d}\eta \mathrm{d}\phi .
\end{equation}

In order to derive the constraints associated to the lifetime of X on $m_\mathrm{DM}$ and $\Delta$, we compute the decay width of X using \texttt{MadGraph5} over a finely grained $(m_\mathrm{DM}, \Delta)$ grid. For each X mass, we approximate $\mathcal{P}_{p_T}$ by generating $p \, p \rightarrow \mathrm{X} \, \overline{\mathrm{X}}$ events. For a large number of generated events $N$, equation~\eqref{eq:distance:longlived} can be reasonably approximated by
\begin{equation}
	P(d_T > d_\mathrm{beam}) = \frac{1}{N} \sum_i \exp \left(-\frac{d_\mathrm{beam}}{d_0^T (p_{Ti})} \right) ,
\end{equation}
where the sum runs over all the events generated. Since the backgrounds for the current LHC searches for long-lived particles are extremely low~\cite{Aad:2013gva,Aad:2015rba,Aaboud:2016dgf,Aaboud:2016uth,Khachatryan:2015jha,Khachatryan:2016sfv}, we consider that a  $(m_{\mathrm{DM}}, \Delta)$ parameter point can be ruled out if at least one long-lived particle is expected to be produced at the working luminosity $\mathcal{L}$. The associated constraint is
\begin{equation}
	2 \times \sigma_{\mathrm{X} \overline{\mathrm{X}}} \times \mathcal{L} \times P(d_T > d_\mathrm{beam}) < 1 ,
\end{equation}
where $\sigma_{\mathrm{X} \overline{\mathrm{X}}}$ is the $\mathrm{X} \, \overline{\mathrm{X}}$ pair-production cross section --- entirely determined by strong interactions --- and the factor of two accounts for the fact that X is pair-produced. These constraints are presented in section~\ref{sec:relic:density} together with the treatment of the relic abundance for our models.

%% file: sections/relic_density_v15.tex
\section{Relic density}
\label{sec:relic:density}
In this section, we present a detailed study of the impact of a strongly interacting coannihilation partner on DM the relic density. As described in section~\ref{sec:dm:partner}, we concentrate on minimal models where the dark matter candidate is a Standard Model singlet with potential new physics couplings smaller than the SM gauge couplings. In spite of its extremely low self-annihilation rates, such a dark matter candidate can easily convert to a strongly interacting coannihilation partner X with whom it is in thermal equilibrium. The partner X can in turn annihilate into quarks and gluons or form $\mathrm{X} \overline{\mathrm{X}}$ bound states, that decay at a later time. Since all these processes occur exclusively through strong interactions, the relic abundance is expected to depend only on the masses of the dark sector particles and the quantum numbers of X. 

In what follows, we derive these relic density bounds for the six models introduced in section~\ref{sec:dm:partner} taking into account non-perturbative effects such as Sommerfeld corrections and bound state formation. For each of these models, we also determine the regions of parameter space for which the DM X SM$_1$ SM$_2$ interaction is large enough to allow the dark matter and X to be in thermal equilibrium.

\subsection{Thermal equilibrium}
\label{sec:thermal:eq}
The efficient depletion of dark matter in our phenomenological scenarios entails chemical and thermal equilibrium with its coannihilation partner X. Establishing this equilibrium requires the existence of DM~$\leftrightarrow$~X exchange processes with a rate larger than the Hubble expansion rate around freeze-out. In our models such processes can take place only through the effective operators presented in equation~\eqref{eq:lagrangians:eft}. Since these operators are suppressed by powers of $\Lambda = 10$~TeV, they are typically associated with low DM~$\leftrightarrow$~X exchange rates. In this subsection, we explicitly compute these rates and comment on the thermal equilibrium constraints on $m_\mathrm{X}$ and $\Delta$ for our models.  

The operators in equation~\eqref{eq:lagrangians:eft} lead to the following three DM~$\leftrightarrow$~X exchange processes
\begin{equation}
	\mathrm{DM} \, \mathrm{SM}_1 \leftrightarrow \mathrm{X} \, \mathrm{SM}_2 \qquad \quad \mathrm{DM} \, \mathrm{SM}_2 \leftrightarrow \mathrm{X} \, \mathrm{SM}_1 \qquad \quad \mathrm{X} \leftrightarrow \mathrm{DM} \, \mathrm{SM}_1 \, \mathrm{SM}_2 \, .
\end{equation}
For thermal equilibrium to take place, the sum of the three corresponding rates in either direction must be larger than the Hubble scale~\cite{Dev:2013yza}, hence
\begin{equation} \label{eq:thermal:eq}
	\Gamma_{\mathrm{DM} \leftrightarrow \mathrm{X}} > \left(\frac{4 \pi^3}{45}\right)^{1/2} g_\rho^{1/2} \frac{m_\mathrm{DM}^2}{x^2 M_{Pl}} \, ,
\end{equation}
where $g_\rho$ is the effective number of relativistic degrees of freedom and $M_{Pl}$ is the Planck mass. The value of $x = m_\mathrm{DM}/T$ at freeze-out has very little model dependence and hence we fix $x_{\mathrm{freeze}} = 25$ to estimate the bound. Note that equation~\eqref{eq:thermal:eq} needs to be satisfied for both the $\mathrm{DM}\rightarrow \mathrm{X}$ and the $\mathrm{X}\rightarrow \mathrm{DM}$ processes. This constraint should therefore be applied on both the $\Gamma_{\mathrm{DM}\rightarrow \mathrm{X}}$ and the $\Gamma_{\mathrm{X}\rightarrow \mathrm{DM}}$ rates, defined as
\begin{equation} \label{eq:thermal:processes}
	\begin{aligned}
		\Gamma_{\mathrm{DM}\rightarrow \mathrm{X}} &= \Gamma_{\mathrm{DM}\,\mathrm{SM}_1\rightarrow \mathrm{X}\,\mathrm{SM}_2} + \Gamma_{\mathrm{DM}\,\mathrm{SM}_2\rightarrow \mathrm{X}\,\mathrm{SM}_1}\\
		\Gamma_{\mathrm{X}\rightarrow \mathrm{DM}} &= \Gamma_{\mathrm{X}\,\mathrm{SM}_1\rightarrow \mathrm{DM}\,\mathrm{SM}_2} + \Gamma_{\mathrm{X}\,\mathrm{SM}_2\rightarrow \mathrm{DM}\,\mathrm{SM}_1} + \Gamma_{\mathrm{X}\rightarrow\mathrm{DM}\,\mathrm{SM}_1\,\mathrm{SM}_2}.
	\end{aligned}
\end{equation}
In the first expression, we have neglected the rate associated with the $\mathrm{DM}\,\mathrm{SM}_1\,\mathrm{SM}_2\rightarrow \mathrm{X}$ process, which is extremely suppressed compared to the other ones.

In order to check the constraint in equation~\eqref{eq:thermal:eq}, we have computed the scattering cross section for the two-to-two processes contributing to equation~\eqref{eq:thermal:processes} for our models. Neglecting the masses of the Standard Model particles, the velocity-averaged exchange rate for processes with a SM fermion in the initial state is~\cite{Ellis:2015vaa}
\begin{equation} \label{eq:thermal:averaged:rate}
	\langle \Gamma_{\mathrm{DM} \leftrightarrow \mathrm{X}} \rangle = \int^\infty_{E_\mathrm{min}} \sigma_{\mathrm{DM} \leftrightarrow \mathrm{X}}(s) \, \frac{g_\mathrm{SM}}{2 \pi^2} \frac{p^2}{e^\frac{p}{T} + 1} \, \mathrm{d} p \, ,
\end{equation}
where $p$ is the momentum and $g_\mathrm{SM}$ the degrees of freedom of the initial state SM particle in the rest frame of the initial state dark sector particle (X or DM) and $E_\mathrm{min}$ is the minimum energy kinematically allowed. For processes with X in the initial state, $E_\mathrm{min} = 0$ while for processes with DM in the initial state $E_\mathrm{min} = (m_\mathrm{X}^2 - m_\mathrm{DM}^2)/(2m_\mathrm{DM})$. If the SM particle in the initial state is a gluon, it needs to obey the Bose-Einstein statistics and equation~\eqref{eq:thermal:averaged:rate} becomes 
\begin{equation} \label{eq:thermal:averaged:rate:boson}
	\langle \Gamma_{\mathrm{DM} \leftrightarrow \mathrm{X}} \rangle = \int^\infty_{E_\mathrm{min}} \sigma_{\mathrm{DM} \leftrightarrow \mathrm{X}}(s) \, \frac{g_\mathrm{SM}}{2 \pi^2} \frac{p^2}{e^\frac{p}{T} - 1} \, \mathrm{d} p \, .
\end{equation}
Finally, the remaining rate $\Gamma_{\mathrm{X}\rightarrow\mathrm{DM}\,\mathrm{SM}_1\,\mathrm{SM}_2}$ is the X decay width and is therefore independent on the velocities of the SM particles involved in the process. We compute it numerically for our different models using~\texttt{MadGraph5}.

For most of the processes considered in this paper, we have verified that the $\mathrm{X}\, \mathrm{SM}_1 \leftrightarrow \mathrm{DM}\, \mathrm{SM}_2$ velocity-averaged rate is several orders of magnitude larger than the Hubble rate at freeze-out in all the regions of parameter space where X decays promptly --- as derived in section~\ref{sec:dm:partner}. The thermal equilibrium constraints are therefore satisfied in all regions of interest for the corresponding models. For the $\mathrm{DM}_\mathrm{S} + \mathrm{X}_\mathrm{F3}$ scenario, however, the thermal equilibrium constraints become significant, especially at large $\Delta$, due to the combination of the loop suppression factor and the $\Lambda^2$ suppression in the effective operator shown in equation~\eqref{eq:lagrangians:eft}. The details of our calculation for this particular scenario are shown in appendix~\ref{sec:thermal:eq:calc}. In the rest of this work, we will therefore show the thermal equilibrium constraints for this model only. These are presented in section~\ref{sec:relic:results} in combination with the lifetime constraints and the thermal relic abundance.

\subsection{Relic density calculation}
\label{sec:relic:density:calculation}
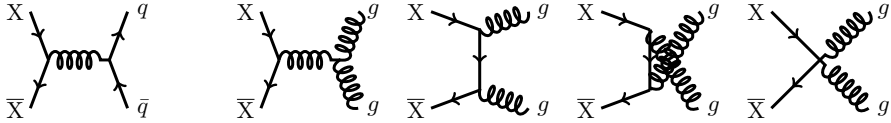
\begin{figure}[!t]
	\centering
	\scalebox{0.8}{\input{diagrams/annihilation_qq_schannel}}
	\hspace{6mm}
	\scalebox{0.8}{\input{diagrams/annihilation_gg_schannel}}
	\hspace{-2mm}
	\scalebox{0.8}{\input{diagrams/annihilation_gg_tchannel}}
	\hspace{-2mm}
	\scalebox{0.8}{\input{diagrams/annihilation_gg_uchannel}}
	\hspace{-2mm}
	\scalebox{0.8}{\input{diagrams/annihilation_gg_4point}}
	\caption{Tree-level processes for the annihilation of X pairs into quark and gluon pairs.}
	\label{fig:treelevel}
\end{figure}

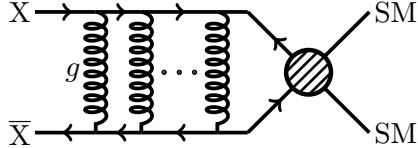
\begin{figure}
	\centering
	\input{diagrams/sommerfeld_ladder}
	\caption{Ladder diagram modeling the Sommerfeld corrections for pair annihilation of X.}
	\label{fig:sommerfeld}
\end{figure}

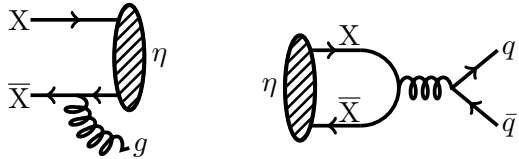
\begin{figure}
	\centering
	\input{diagrams/bound_state_formation}
	\hspace{6mm}
	\raisebox{2.7mm}{\input{diagrams/bound_state_annihilation_decay}}
	\caption{Feynman diagrams for the bound state formation (left) and decay (right) process. The time-reversed bound state formation diagram leads to the dissociation of a bound state via gluon absorption. We assumed that the bound state decays primarily through the annihilation of its two components X and $\overline{\mathrm{X}}$.}
	\label{fig:bsf}
\end{figure}

For each model, we compute the dark matter relic density using \texttt{micrOMEGAs}~\cite{Belanger:2014vza}. As mentioned at the beginning of this section, the dark matter depletion is driven by the $\mathrm{X} \, \overline{\mathrm{X}} \rightarrow q \, \bar{q}$ and $\mathrm{X} \, \overline{\mathrm{X}} \rightarrow g \, g$ processes. The corresponding tree-level interactions are shown in figure~\ref{fig:treelevel}. In addition to these perturbative processes, the X and $\overline{\mathrm{X}}$ initial states also interact through the QCD potential. Since the energy scales considered here are far above the confinement scale, the corresponding interaction can be well described by a Coulomb potential of the form
\begin{equation} \label{eq:potential:qcd}
	V_{\mathrm{QCD}}(r) = -\frac{A}{r}.
\end{equation}
With our definition of the potential, positive $A$ corresponds to an attractive potential while negative $A$ results in a repulsive interaction. The coupling constant $A$ can be written as
\begin{equation} \label{eq:alpha}
	A = \frac{1}{2} (C_\mathrm{X} + C_{\overline{\mathrm{X}}} - C_{\mathrm{X}\overline{\mathrm{X}}}) \alpha_s
\end{equation}
where the $C$ are the quadratic Casimir indices of the X particle or the X$\overline{\mathrm{X}}$ system. This potential describes a long range interaction between the two initial states which can either lead to Sommerfeld corrections~\cite{Sommerfeld:1931aa,deSimone:2014pda,Cassel:2009wt,Cassel:2009pu,Iengo:2009ni} to the tree-level annihilation cross section or to the formation of an X$\overline{\mathrm{X}}$ bound state~\cite{Ellis:2015vaa,Liew:2016hqo,Mitridate:2017izz}. In the former case, the QCD long-range interaction distorts the wave function of the initial X$\overline{\mathrm{X}}$ state, which can lead to sizable modifications of the total annihilation cross section. This non-perturbative phenomenon can be described within a reasonable approximation by the exchange of multiple gluons through ladder diagrams, as shown in figure~\ref{fig:sommerfeld}. Alternatively, X and $\overline{\mathrm{X}}$ can form a bound state and emit a gluon (see figure~\ref{fig:bsf}). This bound state can then either dissociate by reabsorbing a gluon from the thermal bath or decay through the annihilation of its components as shown in figure~\ref{fig:bsf}. At high temperature, when the dissociation dominates over the decay, X$\overline{\mathrm{X}}$ bound state formation does not impact the number density of the dark sector particles. As the temperature decreases, however, the decay width becomes progressively larger, and bound state formation can play a major role in the dark matter annihilation process.

As shown in~\cite{deSimone:2014pda,Ellis:2015vaa,Liew:2016hqo,Harigaya:2014dwa}, the Sommerfeld effect and bound state formation can significantly alter the annihilation cross sections for colored particles, especially at low velocity. In~\cite{ElHedri:2016onc} we described a general and rigorous framework to compute Sommerfeld-corrected annihilation cross sections for models with a single colored coannihilation partner. We also included a \texttt{Mathematica} package~\cite{ElHedri:2016pac} that allows to compute these cross sections for the different annihilation processes and include them in \texttt{micrOMEGAs}. In this work, we improve this package to include effects from the bound state formation and decay, following the procedure described in~\cite{Liew:2016hqo}. The results are detailed in appendix~\ref{sec:bound:states}, where our results obtained for color sextets and vectors are novel. As in~\cite{Liew:2016hqo}, we focus exclusively on $s$-wave color-neutral bound states with zero spin, that are typically associated with the largest formation rates.\footnote{Recently in~\cite{Mitridate:2017izz} an alternative calculation of bound state formation rates for colored particles has been presented, including the effects of higher partial waves, thermal corrections, bound states in the octet representation and the non-Abelian structure of QCD.} Given the inherent uncertainties when dealing with non-perturbative dynamics, the calculated rates should be taken as a mere indication of the expected effect.

\subsection{Results}
\label{sec:relic:results}
\begin{figure}[!t]
	\centering
	\includegraphics[width=0.75\textwidth]{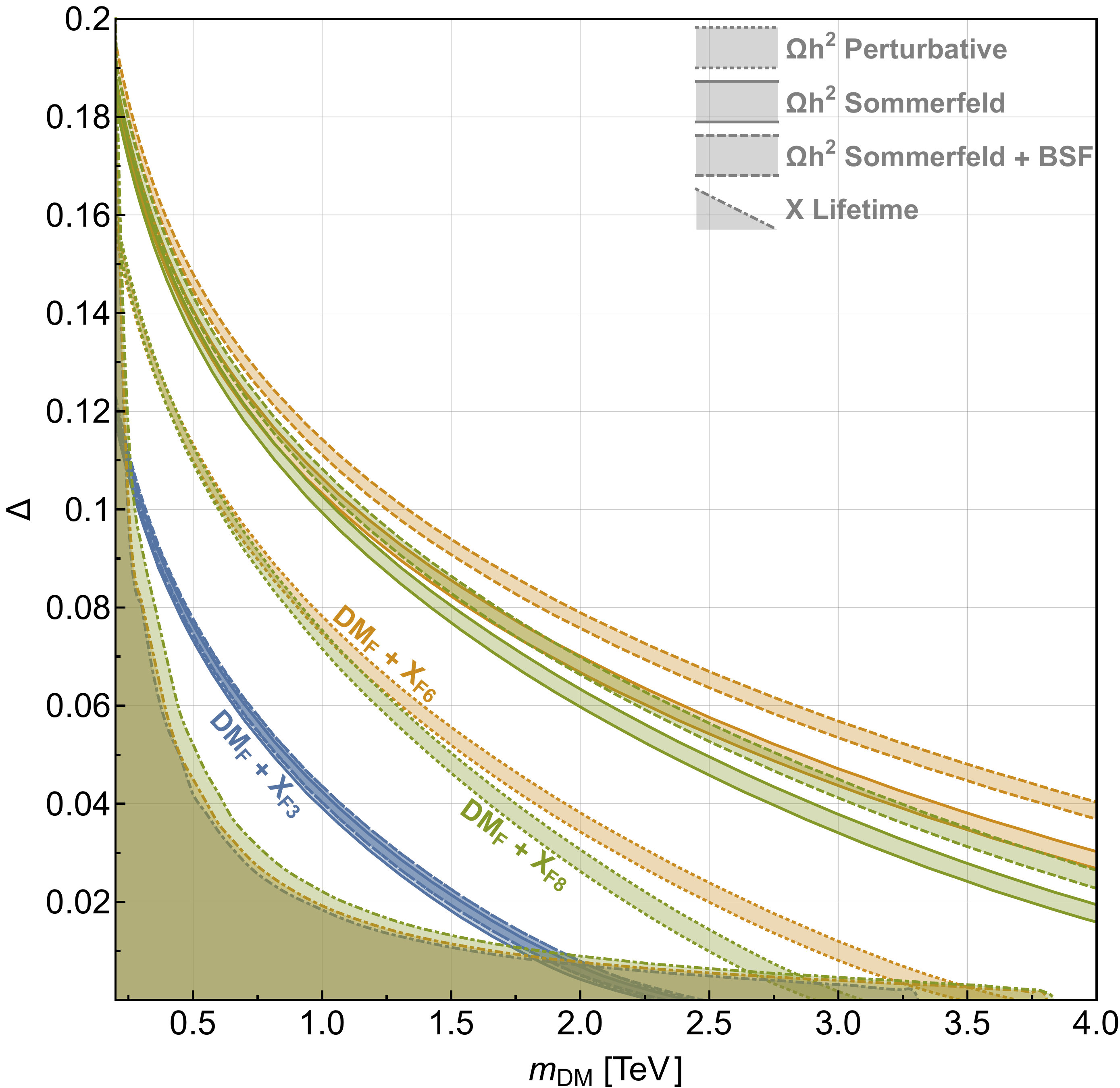}
	\caption{Relic density and lifetime constraints in the $\Delta$ versus $m_\mathrm{DM}$ plane for models where both DM and X are Dirac fermions. We show the parameter space regions that agree with the relic abundance measured by Planck for the  cases of perturbative annihilation only (dotted), with the addition of the Sommerfeld effect (solid) and bound state effects included as well (dashed). The dot-dashed lines enclose the regions that could be potentially excluded by the searches for long-lived particles at LHC13 with $3 \; \mathrm{ab}^{-1}$. In this study, the decay of X is mediated by an effective operator with a suppression scale of $\Lambda = 10$ TeV.}
	\label{fig:relicdensity:mass:delta:set1}
\end{figure}

\begin{figure}[!t]
	\centering
	\includegraphics[width=0.75\textwidth]{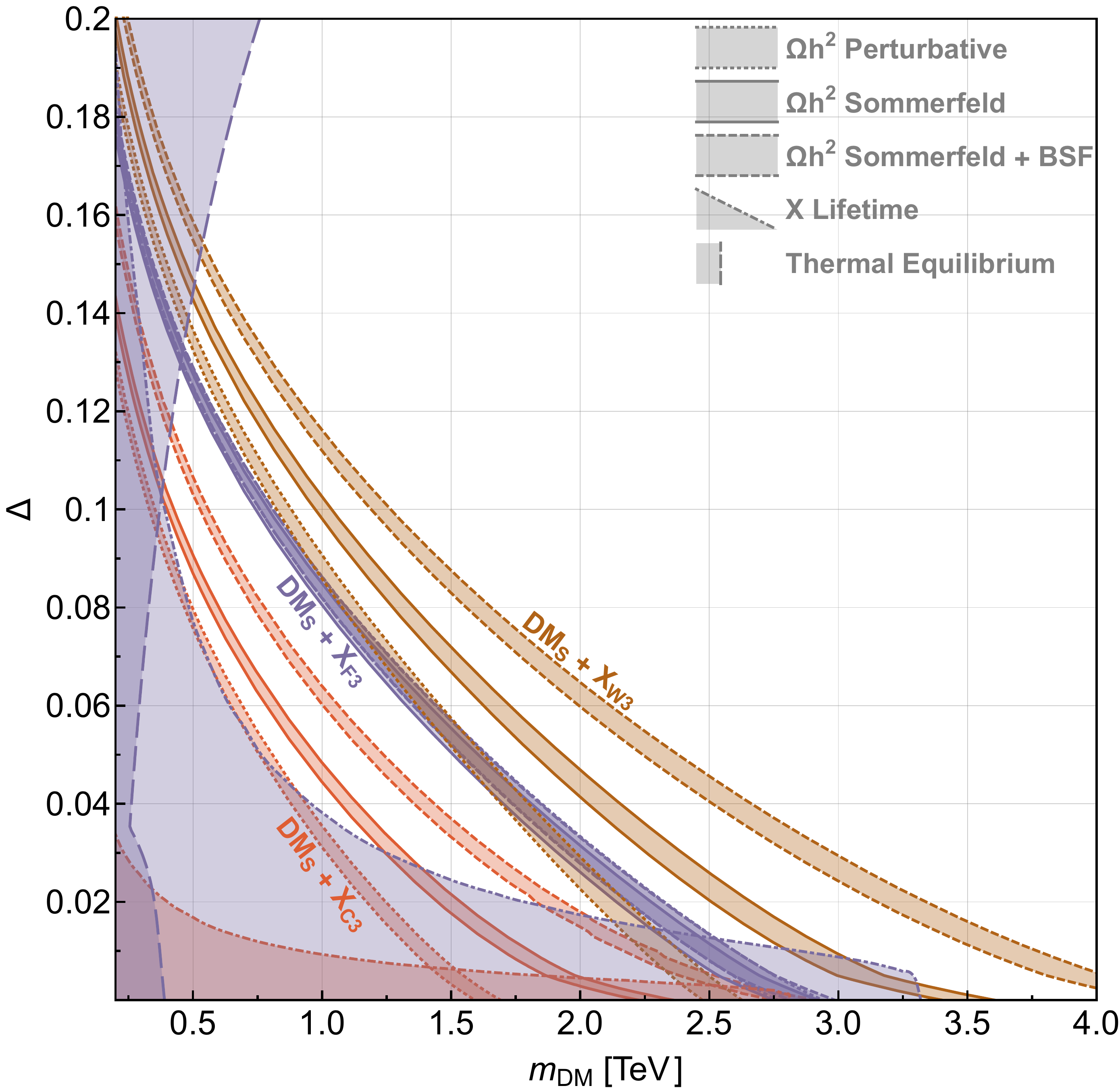}
	\caption{Relic density, lifetime and thermal equilibrium constraints in the $\Delta$ versus $m_\mathrm{DM}$ plane for models where DM is a real scalar and X is a color triplet and either a complex scalar, a Dirac fermion, or a complex vector. We show the parameter space regions that agree with the relic abundance measured by Planck for the  cases of perturbative annihilation only (dotted), with the addition of the Sommerfeld effect (solid) and bound state effects included as well (dashed). The dot-dashed lines enclose the regions that could be potentially excluded by the searches for long-lived particles at LHC13 with $3 \; \mathrm{ab}^{-1}$. In this study, the decay of X and the exchange between DM and X relevant for thermal equilibrium are mediated by an effective operator with a suppression scale of $\Lambda = 10$ TeV. The wide-dashed line shows the thermal equilibrium bound for the $\mathrm{DM}_\mathrm{S} + \mathrm{X}_\mathrm{F3}$ model.}
	\label{fig:relicdensity:mass:delta:set2}
\end{figure}

As discussed in section~\ref{sec:dm:partner} we choose to focus on six representative models among the many possible scenarios with a colored X. The selected models span a wide range of possible spins and color representation of X so that they can be used to estimate the allowed parameter space in any other scenario. The relic density constraints from Planck~\cite{Ade:2015xua} as well as the bounds on the lifetime of X derived in section~\ref{sec:dm:partner} are shown in figures~\ref{fig:relicdensity:mass:delta:set1} and~\ref{fig:relicdensity:mass:delta:set2} in the $\Delta$ versus $m_\mathrm{DM}$ plane. The constraints from the coannihilation partner being long-lived have been calculated assuming LHC13 with an integrated luminosity of $3 \; \mathrm{ab}^{-1}$ as detailed in section~\ref{sec:dm:partner:life}. For each model, we show the annihilation cross sections with and without the Sommerfeld corrections and bound state formation effects.

Figure~\ref{fig:relicdensity:mass:delta:set1} shows the constraints associated with models where both the dark matter and X are Dirac fermions and X is either a color triplet, a sextet or an octet. As expected, the relic density bounds are significantly looser for sextet and octet X than for a triplet. Since the color factors for sextets and octets are similar, the bounds for the corresponding models are of the same order. These models are also both associated with an extremely large Sommerfeld enhancement that extends the allowed range for $m_\mathrm{DM}$ by a factor of $2$ to $3$ for a given $\Delta$. The bound state effects, although significant, are much less pronounced.\footnote{As shown in~\cite{Mitridate:2017izz}, including additional processes could even lead to a much smaller effect.} The inclusion of all these non-perturbative effects pushes the upper bound on dark matter mass to beyond $10$~TeV for X$_\mathrm{F6}$ and X$_\mathrm{F8}$ at low $\Delta$. This bound is undeniably outside the reach of the LHC~\cite{Low:2014cba}, however, it could be within the reach of a future $100$~TeV collider. For X$_\mathrm{F3}$, however, neither the Sommerfeld nor the bound state effect do significantly modify the perturbative cross section. This result arises from an accidental cancellation between the Sommerfeld enhancement for $\mathrm{X} \, \overline{\mathrm{X}} \rightarrow g \, g$ and Sommerfeld depletion for $\mathrm{X} \, \overline{\mathrm{X}} \rightarrow q \, \bar{q}$ channels for models where X is a fermion triplet. As a consequence, the dark matter mass has to be below $2.5$~TeV, which allows a significant fraction of the parameter space to be within the reach of the LHC. It is also important to note that the constraints on these models arising from the lifetime of X are of similar magnitude, and they only intersect the relic abundance bands for the X$_\mathrm{F3}$ model, once the proper corrections have been taken into account.

Figure~\ref{fig:relicdensity:mass:delta:set2} shows the constraints associated with the models where dark matter is a real scalar and where X is a color triplet --- either a complex scalar, a Dirac fermion, or a complex vector. Here as well, the relic density constraints tend to become looser as the number of degrees of freedom of X increases. For the X$_\mathrm{C3}$ and X$_\mathrm{W3}$ models, the bound state and Sommerfeld effects are of about the same order and lead to an order one increase in the dark matter mass for a given $\Delta$. As before, these effects can be neglected for the X$_\mathrm{F3}$ model. The upper bounds on the dark matter mass are between $2.5$ and $3$~TeV for both scalar and fermion X and can be as high as $4.5$~TeV for X$_\mathrm{W3}$. Once again, we observe that the thorough exploration of our models relies on a novel collider with a significantly higher center-of-mass energy than the LHC. Analogously to the case of fermionic X models, the lifetime constraints only lead to a lower bound of about $1$\% on $\Delta$ for the X$_\mathrm{F3}$ and X$_\mathrm{C3}$ models. We note that the lifetime bound for the X$_\mathrm{F3}$ model with scalar DM is stronger than for fermion DM, due to the fact that the corresponding effective operator is suppressed by a loop factor as discussed in section~\ref{sec:dm:partner:life}. Thermal equilibrium constraints supersede the lifetime constraints only for the $\mathrm{DM}_\mathrm{S} + \mathrm{X}_\mathrm{F3}$ model. The "dented" shape of the excluded region is due to the $\Gamma_{\mathrm{X} \rightarrow \mathrm{DM}} > H$ requirement dominating for small $\Delta$, while for large $\Delta$ the $\Gamma_{\mathrm{DM} \rightarrow \mathrm{X}} > H$ condition takes over.

While thermal WIMP dark matter is usually constrained to be lighter than $2-3$~TeV~\cite{Hisano:2006nn,Cirelli:2007xd}, our study shows that the presence of nearby colored states considerably relaxes this bound. In some of these models, dark matter masses up to about $1$~TeV can even be allowed for mass splittings of the order of $10$\% with large X pair-production cross sections. Such a natural region is well within the LHC reach.  For lower values of $\Delta$, however, the allowed values for the dark matter mass become larger, and the need for a more powerful machine becomes evident. While the discussion here about the collider constraints have been kept at a qualitative level, we will present the relevant numerical results for our models in detail in section~\ref{sec:collider:pheno}.

%% file: diagrams/annihilation_qq_schannel.tex
\begin{tikzpicture}[line width=1.4pt, scale=1]
	\draw[fermionbar] (0.8,0.8)--(0.5,0);
	\draw[fermion] (0.8,-0.8)--(0.5,0);
	\draw[fermion] (-0.8,0.8)--(-0.5,0);
	\draw[fermionbar] (-0.8,-0.8)--(-0.5,0);
	\draw[gluon] (-0.5,0)--(0.5,0);
	
	\node at (-1.05,0.8) {$\mathrm{X}$};
	\node at (-1.05,-0.8) {$\overline{\mathrm{X}}$};
	\node at (1.05,0.8) {$q$};
	\node at (1.05,-0.8) {$\bar{q}$};
\end{tikzpicture}

%% file: diagrams/annihilation_gg_schannel.tex
\begin{tikzpicture}[line width=1.4pt, scale=1]
	\draw[gluon] (0.8,0.8)--(0.5,0.0);
	\draw[gluon] (0.8,-0.8)--(0.5,0.0);
	\draw[fermion] (-0.8,0.8)--(-0.5,0);
	\draw[fermionbar] (-0.8,-0.8)--(-0.5,0);
	\draw[gluon] (-0.5,0)--(0.5,0);
	
	\node at (-1.05,0.8) {$\mathrm{X}$};
	\node at (-1.05,-0.8) {$\overline{\mathrm{X}}$};
	\node at (1.05,0.8) {$g$};
	\node at (1.05,-0.8) {$g$};
\end{tikzpicture}

%% file: diagrams/annihilation_gg_tchannel.tex
\begin{tikzpicture}[line width=1.4pt, scale=1]
	\draw[gluon] (0.8,0.8)--(0,0.5);
	\draw[gluon] (0.8,-0.8)--(0,-0.5);
	\draw[fermion] (-0.8,0.8)--(0,0.5);
	\draw[fermionbar] (-0.8,-0.8)--(0,-0.5);
	\draw[fermion] (0,0.5)--(0,-0.5);
	
	\node at (-1.05,0.8) {$\mathrm{X}$};
	\node at (-1.05,-0.8) {$\overline{\mathrm{X}}$};
	\node at (1.05,0.8) {$g$};
	\node at (1.05,-0.8) {$g$};
\end{tikzpicture}

%% file: diagrams/annihilation_gg_uchannel.tex
\begin{tikzpicture}[line width=1.4pt, scale=1]
	\draw[gluon] (0.8,0.8)--(0,-0.5);
	\draw[gluon] (0.8,-0.8)--(0,0.5);
	\draw[fermion] (-0.8,0.8)--(0,0.5);
	\draw[fermionbar] (-0.8,-0.8)--(0,-0.5);
	\draw[fermion] (0,0.5)--(0,-0.5);
	
	\node at (-1.05,0.8) {$\mathrm{X}$};
	\node at (-1.05,-0.8) {$\overline{\mathrm{X}}$};
	\node at (1.05,0.8) {$g$};
	\node at (1.05,-0.8) {$g$};
\end{tikzpicture}

%% file: diagrams/annihilation_gg_4point.tex
\begin{tikzpicture}[line width=1.4pt, scale=1]
	\draw[gluon] (0.8,0.8)--(0.0,0.0);
	\draw[gluon] (0.8,-0.8)--(0.0,0.0);
	\draw[fermion] (-0.8,0.8)--(0,0);
	\draw[fermionbar] (-0.8,-0.8)--(0,0);
	
	\node at (-1.05,0.8) {$\mathrm{X}$};
	\node at (-1.05,-0.8) {$\overline{\mathrm{X}}$};
	\node at (1.05,0.8) {$g$};
	\node at (1.05,-0.8) {$g$};
\end{tikzpicture}

%% file: diagrams/sommerfeld_ladder.tex
\begin{tikzpicture}[line width=1.4pt, scale=1]
	\draw[fermionbar] (-2.8,0.8)--(-3.6,0.8);
	\draw[fermion] (-2.8,-0.8)--(-3.6,-0.8);
	\draw[fermionbar] (-2.2,0.8)--(-2.8,0.8);
	\draw[fermion] (-2.2,-0.8)--(-2.8,-0.8);
	\draw[fermionbar] (-1.2,0.8)--(-2.2,0.8);
	\draw[fermion] (-1.2,-0.8)--(-2.2,-0.8);
	\draw[fermionna] (-1.2,0.8)--(-0.8,0.8);
	\draw[fermionna] (-1.2,-0.8)--(-0.8,-0.8);
	\draw[gluon] (-2.8,0.8)--(-2.8,-0.8);	
	\draw[gluon] (-2.2,0.8)--(-2.2,-0.8);
	\draw[gluon] (-1.2,0.8)--(-1.2,-0.8);
	\draw[fermionna] (0.8,0.8)--(0,0);
	\draw[fermionna] (0.8,-0.8)--(0,0);
	\draw[fermionbar] (-0.8,0.8)--(0,0);
	\draw[fermion] (-0.8,-0.8)--(0,0);
	\draw[fill=black] (0,0) circle (3.0mm);
	\draw[fill=white] (0,0) circle (2.9mm);
	\begin{scope}
		\clip (0,0) circle (3.0mm);
		\foreach \x in {-.9,-.75,...,.3}
		\draw[line width=1 pt] (\x,-.3) -- (\x+.6,.3);
	\end{scope}
	
	\draw[fill=black] (-1.5,0) circle (0.1mm);
	\draw[fill=black] (-1.7,0) circle (0.1mm);
	\draw[fill=black] (-1.9,0) circle (0.1mm);
	
	\node at (-3.8,0.8) {$\mathrm{X}$};
	\node at (-3.1,0.0) {$g$};
	\node at (-3.8,-0.8) {$\overline{\mathrm{X}}$};
	\node at (1.15,0.8) {SM};
	\node at (1.15,-0.8) {SM};
\end{tikzpicture}

%% file: diagrams/bound_state_formation.tex
\begin{tikzpicture}[line width=1.4pt, scale=1]
	\draw[fermion] (-0.3,0.5)--(0.9,0.5);
	\draw[fermionbar] (-0.3,-0.5)--(0.3,-0.5);
	\draw[fermionbar] (0.3,-0.5)--(0.9,-0.5);
	\draw[fill=black] (1,0) ellipse (2.0mm and 7.0mm);
	\draw[fill=white] (1,0) ellipse (1.95mm and 6.93mm);
	\begin{scope}
		\clip (1,0) ellipse (2.0mm and 7.0mm);
		\foreach \x in {-3.9,-3.75,...,1.9}
		\draw[line width=1pt] (\x,-1.5) -- (\x+3.0,1.5);
	\end{scope}
	\draw[gluon] (0.3,-0.5) arc (0:90:-0.7);
	\node at (1.4,0.0) {$\eta$};
	\node at (-0.45,0.5) {$\mathrm{X}$};
	\node at (-0.45,-0.5) {$\overline{\mathrm{X}}$};
	\node at (1.15,-1.2) {$g$};
\end{tikzpicture}

%% file: diagrams/bound_state_annihilation_decay.tex
\begin{tikzpicture}[line width=1.4pt, scale=1]
	\draw[fermion] (-0.9,0.5)--(-0.2,0.5);
	\draw[fermionbar] (-0.9,-0.5)--(-0.2,-0.5);	
	\draw[fill=black] (-1,0) ellipse (2.0mm and 7.0mm);
	\draw[fill=white] (-1,0) ellipse (1.95mm and 6.93mm);
	\begin{scope}
		\clip (-1,0) ellipse (2.0mm and 7.0mm);
		\foreach \x in {-3.9,-3.75,...,1.9}
		\draw[line width=1pt] (\x,-1.5) -- (\x+3.0,1.5);
	\end{scope}
	\draw[fermionna] (-0.2,-0.5) arc (-90:90:0.5);
	\draw[gluon] (0.3,0.0)--(1.0,0.0);
	\draw[fermion] (1.0,0.0)--(1.6,0.5);
	\draw[fermionbar] (1.0,0.0)--(1.6,-0.5);
	\node at (-1.4,0.0) {$\eta$};
	\node at (-0.35,0.7) {$\mathrm{X}$};
	\node at (-0.35,-0.28) {$\overline{\mathrm{X}}$};
	\node at (1.75,0.5) {$q$};
	\node at (1.75,-0.5) {$\bar{q}$};
\end{tikzpicture}

%% file: sections/collider_v15.tex
\section{Collider phenomenology}
\label{sec:collider:pheno}
The models considered in this study all share the same collider phenomenology. The collider signatures of simplified models of coannihilating dark matter have been classified in~\cite{Baker:2015qna} for all possible choices of quantum numbers of X. In reference~\cite{Buschmann:2016hkc} we explored the phenomenology of dark matter models where X is colored and the coannihilation process occurs through an $s$-channel mediator. These scenarios lead to a wide variety of characteristic collider signatures, notably from mediator single and pair-production. When the mediator becomes heavy, however, the most striking signature for coannihilation arises from the pair-production of X in association with jets and its subsequent $\mathrm{X} \rightarrow \mathrm{DM} \, j \, j$ decay. This process leads to signatures with jets plus missing transverse energy that are already being probed by the current ATLAS and CMS searches~\cite{Aaboud:2016tnv,ATLAS-CONF-2016-078,Aaboud:2016zdn,CMS-PAS-SUS-16-016,CMS-PAS-SUS-16-014}. These signatures are universally shared between all models of coannihilation with a strongly coupled coannihilation partner.

Signatures with jets plus MET are currently being targeted by the monojet searches for dark matter particles and by the multijet plus MET searches tailored for squarks and gluinos. The monojet searches look for signatures with at least one hard jet ($p_T > 500$~GeV) recoiling against large missing energy, allowing for --- but not cutting on --- additional softer jets. These searches are particularly sensitive to events where either an invisible particle is pair-produced, or the visible decay products of a particle are too soft to be used in a search. They are therefore particularly suited for investigating our coannihilation models in the region where the mass splitting $\Delta$ between the dark matter and X is small. As $\Delta$ increases, however, the jets coming from the X decay become harder and the corresponding signature becomes increasingly similar to the ones probed by the ATLAS and CMS multijet searches. These searches are in fact targeting the exact same $\mathrm{X} \rightarrow \mathrm{DM} \, j \, j$ decay process as the one studied here but are primarily focusing on regions of parameter space with a sizable splitting between DM and X.

The goal of this section is to determine how the current and future LHC results for monojet and multijet searches constrain our models. We first review the details of the current ATLAS and CMS analyses, and derive the corresponding exclusion bounds in terms of the \emph{observed} $95$\% CL limits . We then extrapolate the current \emph{expected} $95$\% confidence limits in order to obtain future projections for the high-luminosity LHC with 3000 fb$^{-1}$  of total integrated luminosity, paying particular attention to the role of systematics. Finally we compare the LHC bounds to the relic-density favored region of parameter space to determine the ultimate reach of the LHC for models that lead to the observed dark matter relic density. This information is crucial in order to design effective probes of these models at a putative future collider with higher center-of-mass energy.

\subsection{LHC searches}
\label{sec:lhc:searches}
In this study, we consider the existing jets + MET searches from ATLAS and CMS. Both collaborations select events with a large missing energy ($\geq 100$~GeV at the trigger level, $\geq 200$~GeV in the pre-selection stage) and no reconstructed leptons. ATLAS presented a monojet analysis~\cite{Aaboud:2016tnv} (including up to 4 jets) using a total luminosity of $3.2$~fb$^{-1}$ and a multi-jet analysis including up to 6 jets and using a $13.3$~fb$^{-1}$ dataset~\cite{ATLAS-CONF-2016-078}, which supersedes the $3.2$~fb$^{-1}$ study~\cite{Aaboud:2016zdn}. In addition there is a search considering jet multiplicities between 8 and 10~\cite{ATLAS-CONF-2016-095}, but since we expect the number of jets in our signals to be much lower, this analysis will not be included in this work.
 
CMS has carried out similar studies~\cite{CMS-PAS-SUS-16-016,CMS-PAS-SUS-16-014}, where different bins in the number of jets, the number of b-jets and additional variables are considered, giving 72 signal regions in their monojet analysis and 160 regions in their multi-jet analysis, both carried out at $12.9$~fb$^{-1}$. Since the CMS collaboration has not presented the final numbers of background events in each signal region, but only a log-scaled histogram, we decided to use only the ATLAS data.\footnote{A fairer comparison can be done between the ATLAS studies and the previous CMS analysis~\cite{Khachatryan:2016kdk,CMS-PAS-SUS-15-005} using $2.3$~fb$^{-1}$ of data, where comparable bounds were obtained when interpreting the data under the same hypothesis, for example a $500$--$600$~GeV lower limit on squark masses, depending on the mass gap to the neutralino.}

\subsubsection{Monojet ATLAS search}
The ATLAS monojet study~\cite{Aaboud:2016tnv} selects events with  $\met > 250$~GeV as well as a leading jet with $p_T(j_1) > 250$~GeV and $|\eta(j_1)| < 2.4$. Up to three additional jets with $p_T > 30$~GeV and $| \eta| < 2.8$ are allowed but not used in the search. These jets need to satisfy $\Delta \phi(j,\vec{p}_T^{\rm miss}) > 0.4$ in order to reject the QCD multi-jet background arising from mismeasuring the jet momenta. In addition, electrons (muons) with $p_T > 10$~($20$)~GeV are vetoed. Inclusive signal regions for $\met > x=250, 300, \dots, 700$~GeV (dubbed IM1--IM7) and exclusive signal regions for $\met \in [x, x + 50]$~GeV (EM1--EM6) are defined. The corresponding $95$\% CL upper limits on number of signal events and cross sections is then computed for each signal region separately.

\subsubsection{Multi-jet ATLAS search}
The ATLAS multi-jet study~\cite{ATLAS-CONF-2016-078} defines signal regions according to the number of jets, ranging from 2 to 6, and to the lower value chosen for the effective mass $m_{\rm eff}$, which is strongly correlated with the degree of background rejection.\footnote{This analysis uses the effective mass with the leading $N_j$ jets, $m_{\rm eff} (N_j)$ for the $N_j$ bin, as well as the 'inclusive' one where the sum is done over all existing jets with $p_T > 50$ GeV. For simplicity we refer to the latter as $m_{\rm eff}$.} The baseline requirements are $\met > 250~\rm{GeV}$ and the absence of leptons. Each signal region has its own thresholds for the transverse momenta of the jets, the minimum azimuthal separation between the jets and the missing energy, $\Delta \phi(j,\vec{p}_T^{\rm miss})_{\rm min}$ and on $\met / m_{\rm eff} (N_j)$ or $\met / \sqrt{H_T}$. Additional cuts on the pseudorapidity differences between the jets and/or on the so-called ``aplanarity'' variable~\cite{Chen:2011ah} apply in certain regions. We can establish a clear distinction between the regions where two or more hard jets are requested and those where only one hard jet is requested and the additional jets are not vetoed, which we dub monojet-like. Our naive expectation is that, due to the low $\Delta$ splitting, our models would be constrained mostly by the monojet-like signal regions from the multi-jet search. As for the monojet search, the $95$\% confidence limits on the number of signal events are estimated for each signal region. Note that this multi-jet analysis also presents a new technique called the "recursive jigsaw reconstruction" , aimed at compressed gluinos. Although this search is expected to give better results, there is not enough information given to be able to recast it. We should therefore keep in mind that our bounds are conservative.

\subsection{Recasting and extrapolation}
\label{sec:lhc:recasting}
In order to recast the current exclusion bounds into our models, we simply use the $95$\% CL \emph{observed} upper limits on the number of signal events,  $S^{\rm 95}_{\rm obs}$ provided in~\cite{Aaboud:2016tnv} and~\cite{ATLAS-CONF-2016-078}. We take these limits to be equivalent to a significance of two standard deviations of a Gaussian distribution (2$\sigma$). Thus the significance for a given signal region is simply $2 S_i/ S^{\rm 95}_{\rm obs}$, where $S_i$ is the number of signal events expected in the signal region under consideration for our model.

To extrapolate the current LHC results to higher luminosities, we rely on the \emph{expected} $95$\% CL upper limit on the number of signal events $S^{\rm 95}_{\rm exp}$ which we take as
\begin{equation} \label{eq:sigest}
	S^{\rm 95}_{\rm exp} \simeq 2 \, \delta \! B =  2 \sqrt{B_i + \beta_i^2 B_i^2} ,
\end{equation}
where $\beta_i$ is the systematic uncertainty. This equation is derived by assuming that the expected fluctuation in the number of background events $\delta \! B$ has a statistical and a systematic component. Since the correlation between both uncertainties is not reported by the ATLAS collaboration, we assume no correlation and then combine them in quadrature. The significance $S_i/ \delta \! B$ recovers the well known limit of $S_i / \sqrt{B_i}$, which scales as the square root of the luminosity, in the absence of systematics. When the statistical errors become negligible, however, the significance can be approximated by $S_i / (\beta_i B_i)$ and no longer depends on the luminosity.

In view of the previous discussion, it is crucial to establish a procedure to accurately estimate the systematic uncertainties, and to validate it using the available data. We first consider the monojet analysis, where the main background contribution arises from $Z ( \to \nu \nu )$~+~jets. To first approximation, we consider only this background and use equation~\eqref{eq:sigest} to determine the value of $\beta$ that allows to reproduce the value of $S^{\rm 95}_{\rm exp}$ given in~\cite{Aaboud:2016tnv} for each signal region. In IM1, we find $\beta = 5.5$\%. When repeating this procedure using the sum of the background contributions, this number moves to $4.3$\%. These values are compatible with the $2$--$4$\% total background uncertainties given in~\cite{Aaboud:2016tnv} that take possible correlations between the statistic and systematic errors into account. For the EM3, EM5 and EM7 signal regions, the values of $\beta$ obtained using the $Z$~+~jets (total) background are of $6.3$\% $(5.0\%)$, $8.2$\% $(6.2\%)$ and $13.5$\% $(9.1\%)$ respectively. Hence we can safely assume that the overall systematic uncertainty on the background is dominated by the $Z ( \to \nu \nu )$~+~jets contribution. This approximation also allows for a conservative estimate of the systematic error throughout all signal regions. In each region, we find this systematic error to be the dominant source of background uncertainties, $\beta \, \sqrt{B} > 1$.

\begin{figure}[!t]
	\centering
	\includegraphics[width=0.8\textwidth]{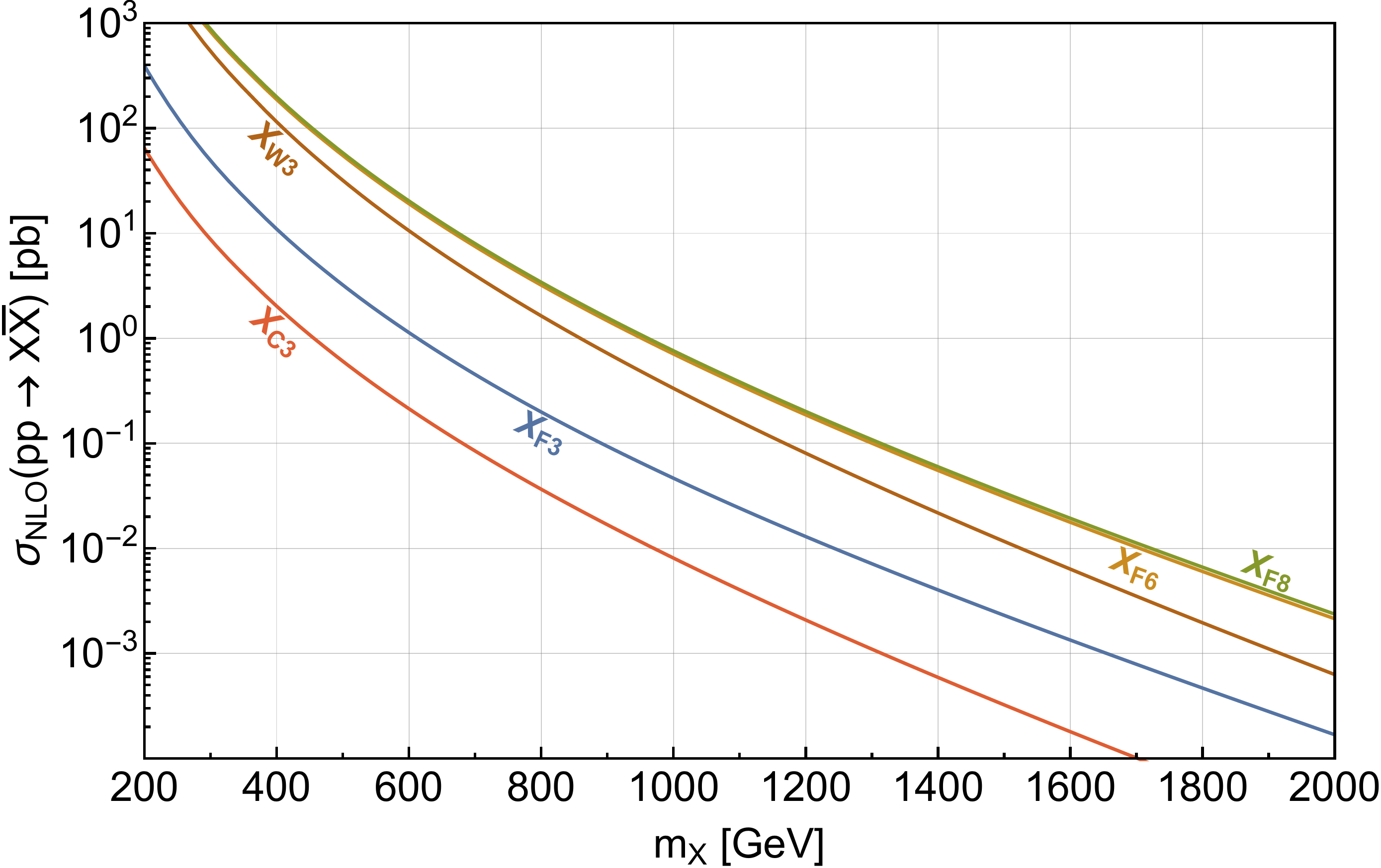}
	\caption{NLO production cross sections for $p \, p \to \mathrm{X} \, \overline{\mathrm{X}}$ for our models using mass dependent K-factors.}
	\label{fig:xsplot}
\end{figure}

In the multi-jet analysis~\cite{ATLAS-CONF-2016-078}, the sub-leading jets are required to be hard in most of the signal regions, which cuts deeper into phase space, giving rise to larger statistical uncertainties compared to the monojet case. Here, we estimate the systematic uncertainties using the same procedure as for the monojet searches, this time using the sum of all the backgrounds for each signal region. These uncertainties now range from 8\% in the '2j-800' bin to $23$\% in the '4j-2200' bin, while the '6j-2200' bin has a systematic error of $43$\%. Although these values are considerably larger than for the monojet search, the number of background events in the signal regions is now extremely low and the statistical uncertainties contribute at least as much as the systematic uncertainties to the global error everywhere except in the 2-jet bins.\footnote{In several signal regions in~\cite{ATLAS-CONF-2016-078}, the quoted error on the background is less than $\sqrt{B}$. This is because, as mentioned in the search, this number does not take the statistical error into account.} Having in mind the high-luminosity phase, we stress that with the recently collected dataset of about 40 fb$^{-1}$ the statistical error will decrease to a point where the systematic uncertainties will become relevant again.

\subsection{Results}
\label{sec:lhc:results}
In this section, we present the overall bounds on the selection of models presented in section~\ref{sec:dm:partner} from the recasted ATLAS monojet and multijet analyses described previously. For each model, we scan the parameter space over ranges informed by the relic density constraints from section~\ref{sec:relic:density}, namely for $m_\mathrm{DM} \in [200, 2000]$~GeV and $\Delta \in [0,0.2]$. We simulated the signal events using \texttt{MadGraph5}~\cite{Alwall:2014hca} with the \texttt{CTEQ6L1} parton distribution functions~\cite{Pumplin:2002vw}, interfaced with \texttt{Pythia v8.2}~\cite{Sjostrand:2014zea} for parton showering and hadronization. The signal events are matched up to two jets using the MLM procedure with the $k_\perp$-showering scheme~\cite{Hoche:2006ph,Alwall:2007fs,Alwall:2008qv} since, for small values of $\Delta$, a proper description of the sub-leading jets is necessary in order to accurately use the current experimental results.  Basic detector simulation is performed in \texttt{Delphes v3.3.3}~\cite{deFavereau:2013fsa}. The parton level cross sections (prior to matching) for $p \, p \to \mathrm{X} \, \overline{\mathrm{X}}$ at next-to-leading order (NLO) are shown in figure~\ref{fig:xsplot}. We obtain the NLO result by multiplying the leading-order (LO) cross section obtained with \texttt{MadGraph5} by a mass-dependent K-factor for each model. The K-factors are computed at NLO using \texttt{Top++ v2.0}~\cite{Czakon:2011xx} interfaced with \texttt{LHAPDF v6}~\cite{Buckley:2014ana} and \texttt{CTEQ6L1}~\cite{Pumplin:2002vw} for the X$_\mathrm{F3}$ models and \texttt{Prospino v2.1}~\cite{Beenakker:1996ch} for the X$_\mathrm{C3}$ and X$_\mathrm{F8}$ models. The K-factors of these last two models are identical to the ones for squark-antisquark production and gluino pair-production respectively. To date, the K-factors for pair-production of fermion sextets and vector triplets have not been computed. We therefore take the K-factors for X$_\mathrm{F6}$  and X$_\mathrm{W3}$ to be equal to the K-factors of X$_\mathrm{F8}$ and X$_\mathrm{C3}$ respectively. The similarity of the $\mathrm{X}_\mathrm{F8}$ and $\mathrm{X}_\mathrm{F6}$ cross sections is due to the similar values of the quadratic Casimir indices for the sextet and octet color representations. Hence, we expect the mass reach for these two models to be similar. We observe that our representative set of colored dark sector models span two orders of magnitude in cross section, with the $\mathrm{X}_\mathrm{C3}$ model giving the lowest values (complex scalar, color triplet) while the highest values are obtained for $\mathrm{X}$ being a fermion and either a sextet or an octet of $SU(3)$. 
 
\begin{figure}[!t]
	\centering
	\includegraphics[width=0.49\textwidth]{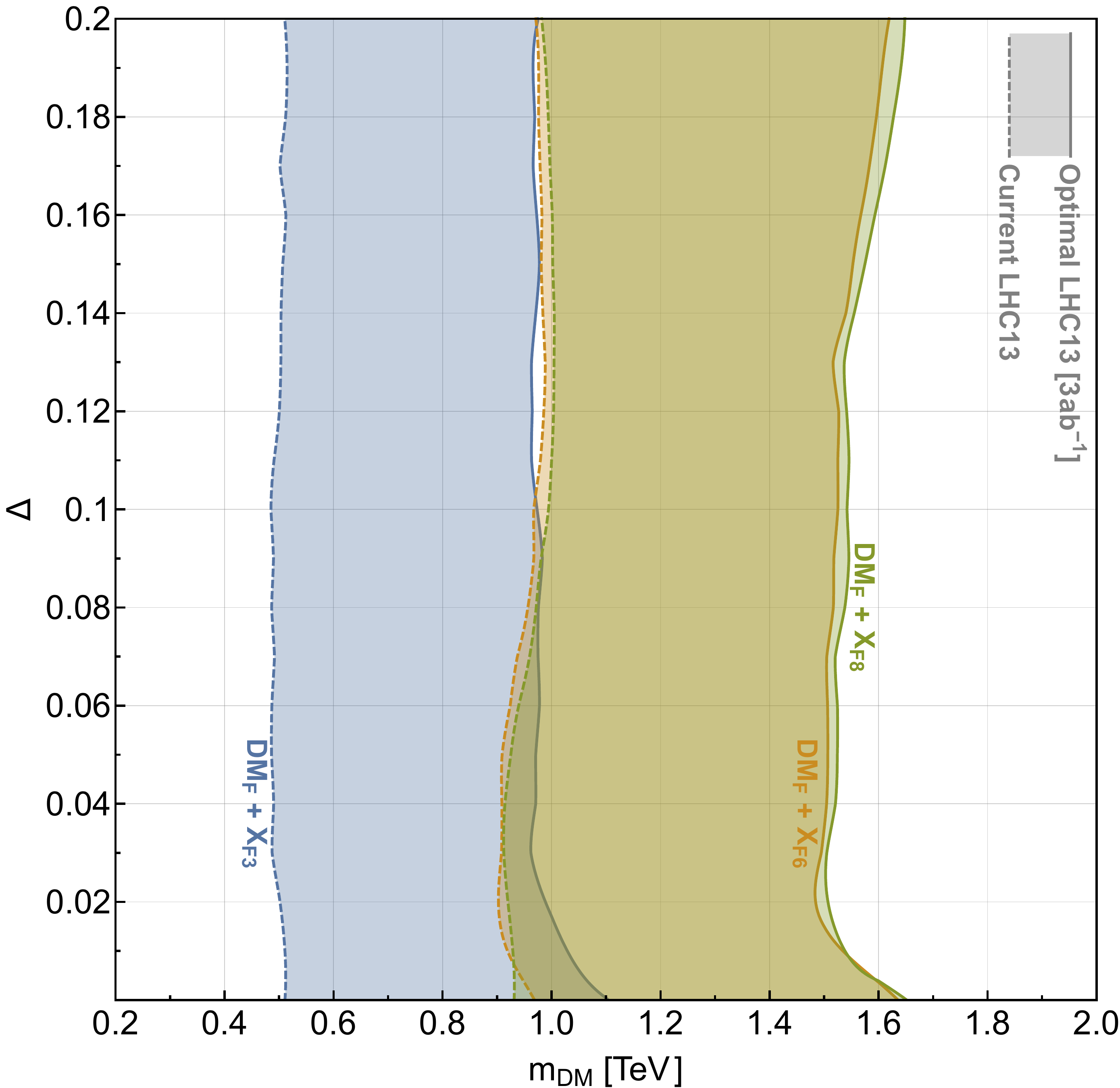}
	\includegraphics[width=0.49\textwidth]{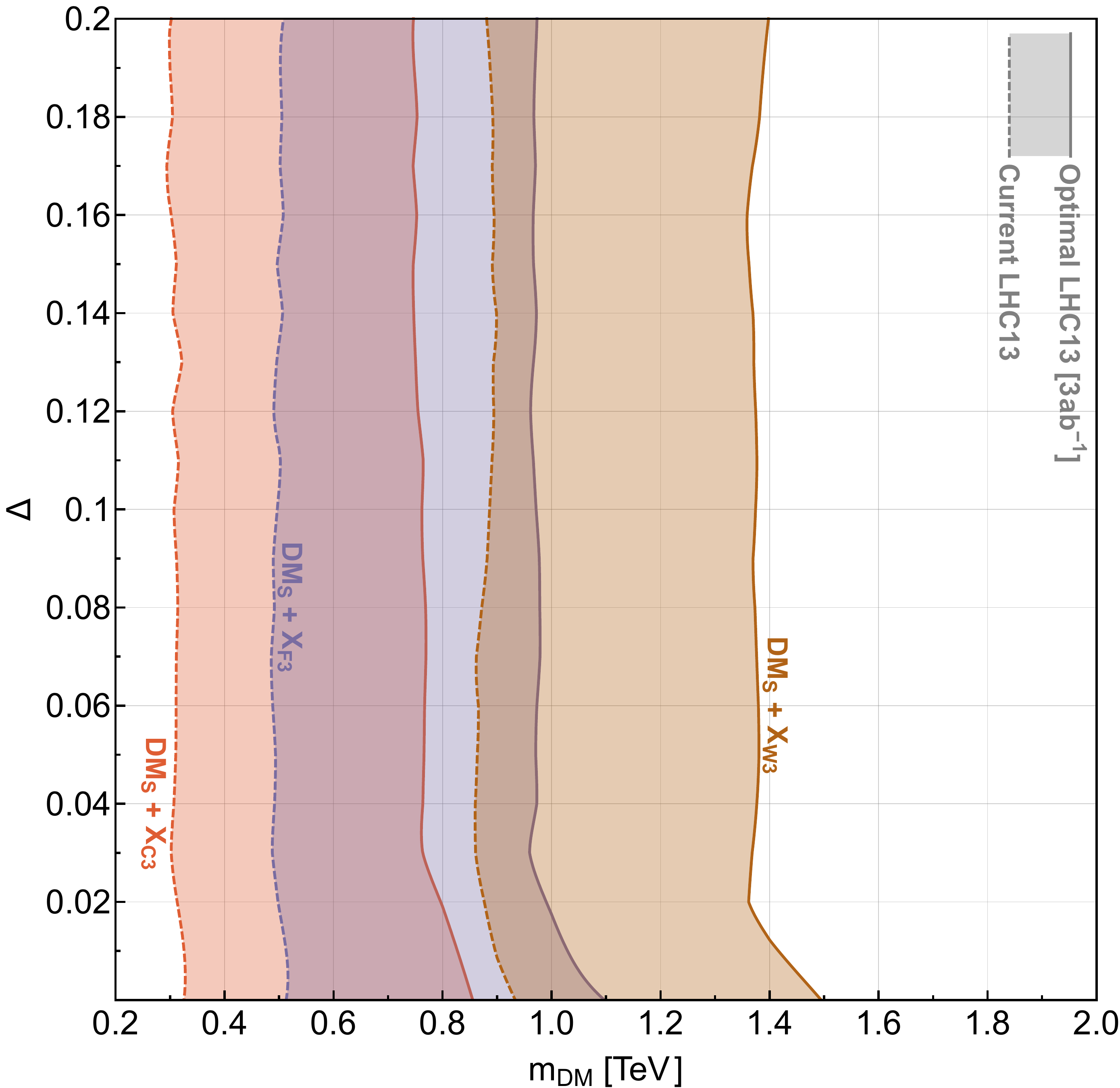}
	\caption{$95$\% CL exclusion bounds on our sample models. The bands range from the current exclusions (dashed lines) to the expected exclusions with $3000$ fb$^{-1}$ while neglecting systematic effects (solid lines). The left panel shows these bounds for all models where the coannihilating partner X is a fermion while the right panel shows the bounds for all models where X is a color triplet.}
	\label{fig:lhc:exclusion:bands}
\end{figure}

\begin{figure}[!t]
	\centering
	\includegraphics[width=0.48\textwidth]{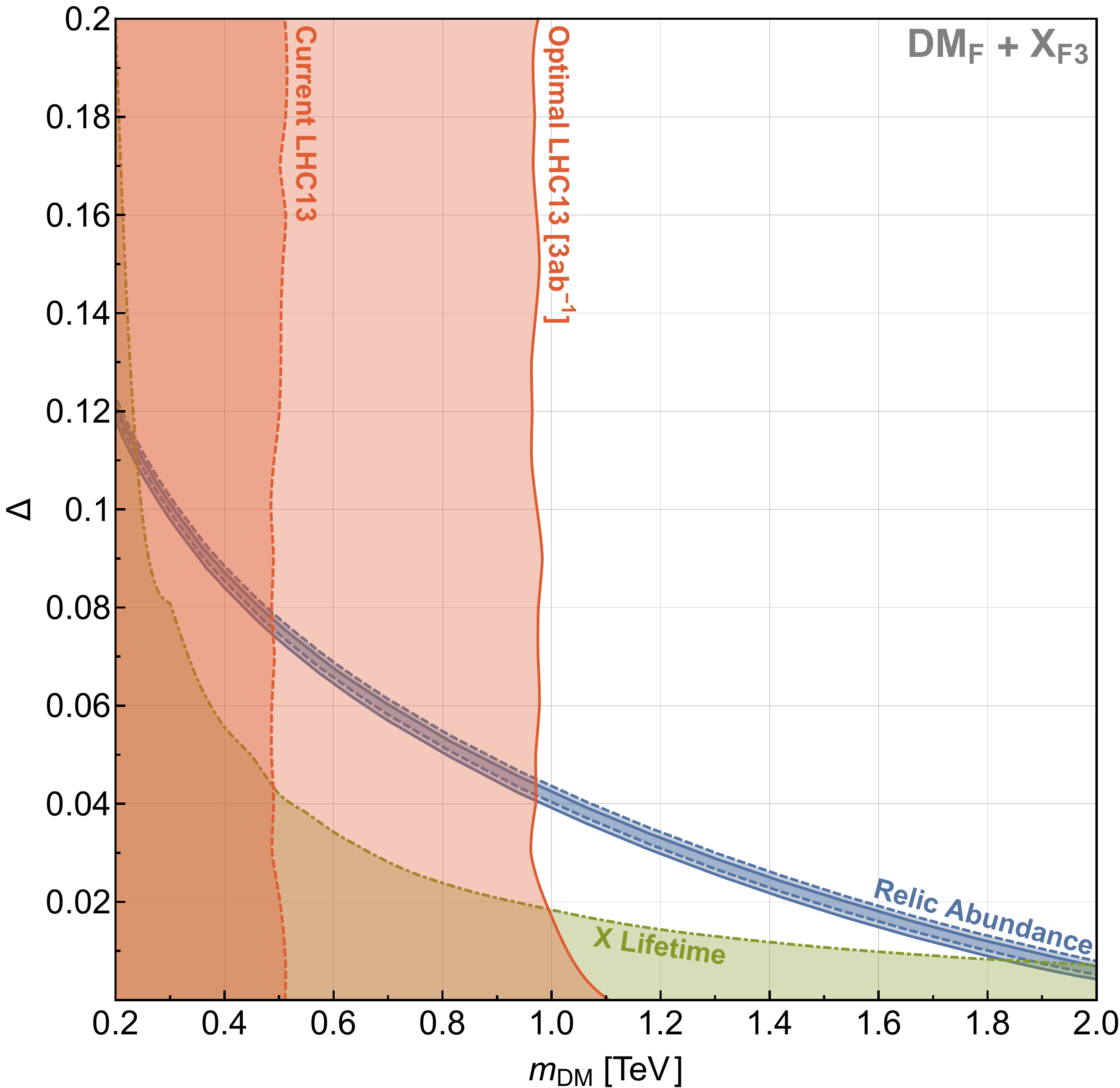}
	\includegraphics[width=0.48\textwidth]{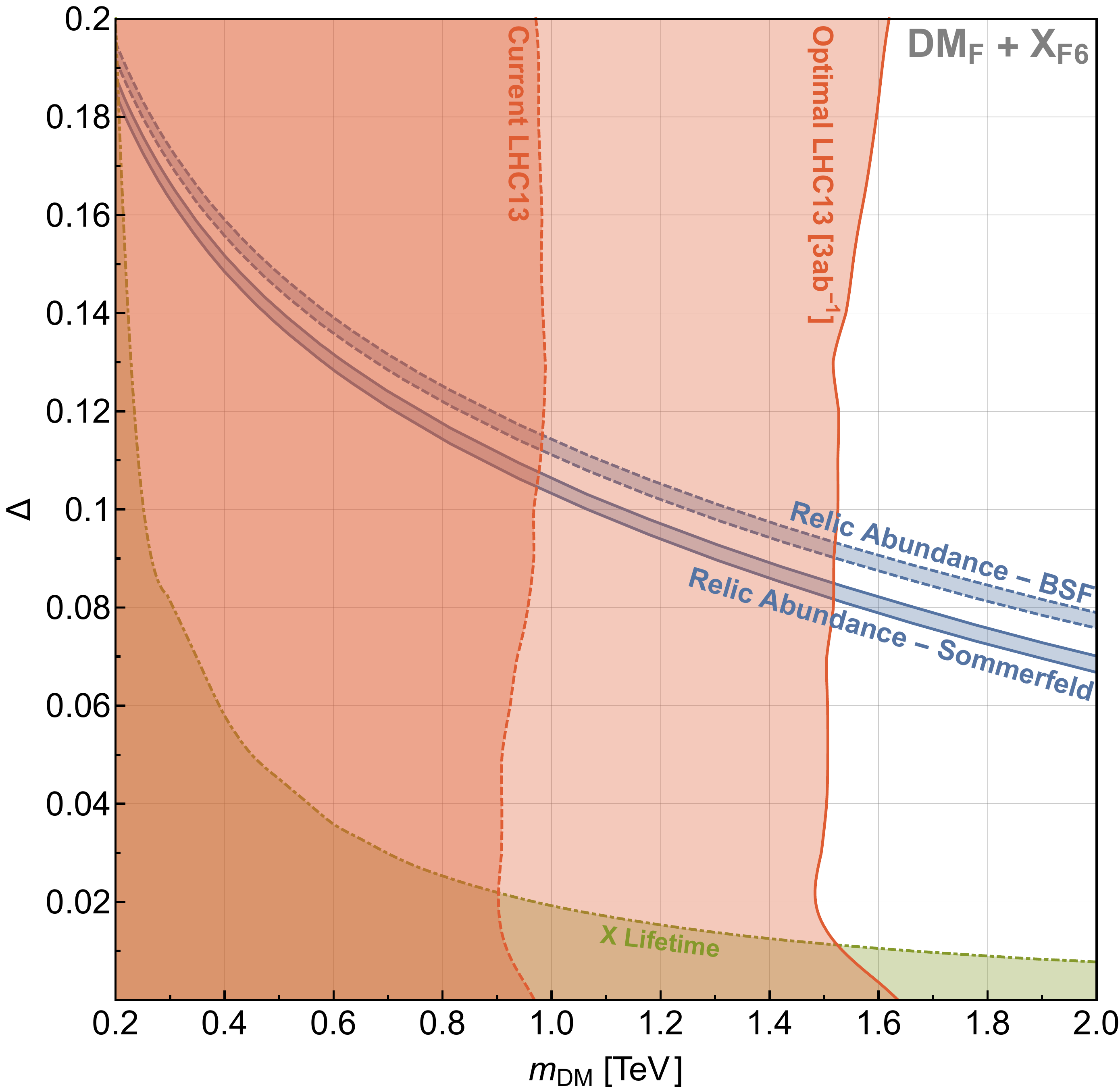}
	\includegraphics[width=0.48\textwidth]{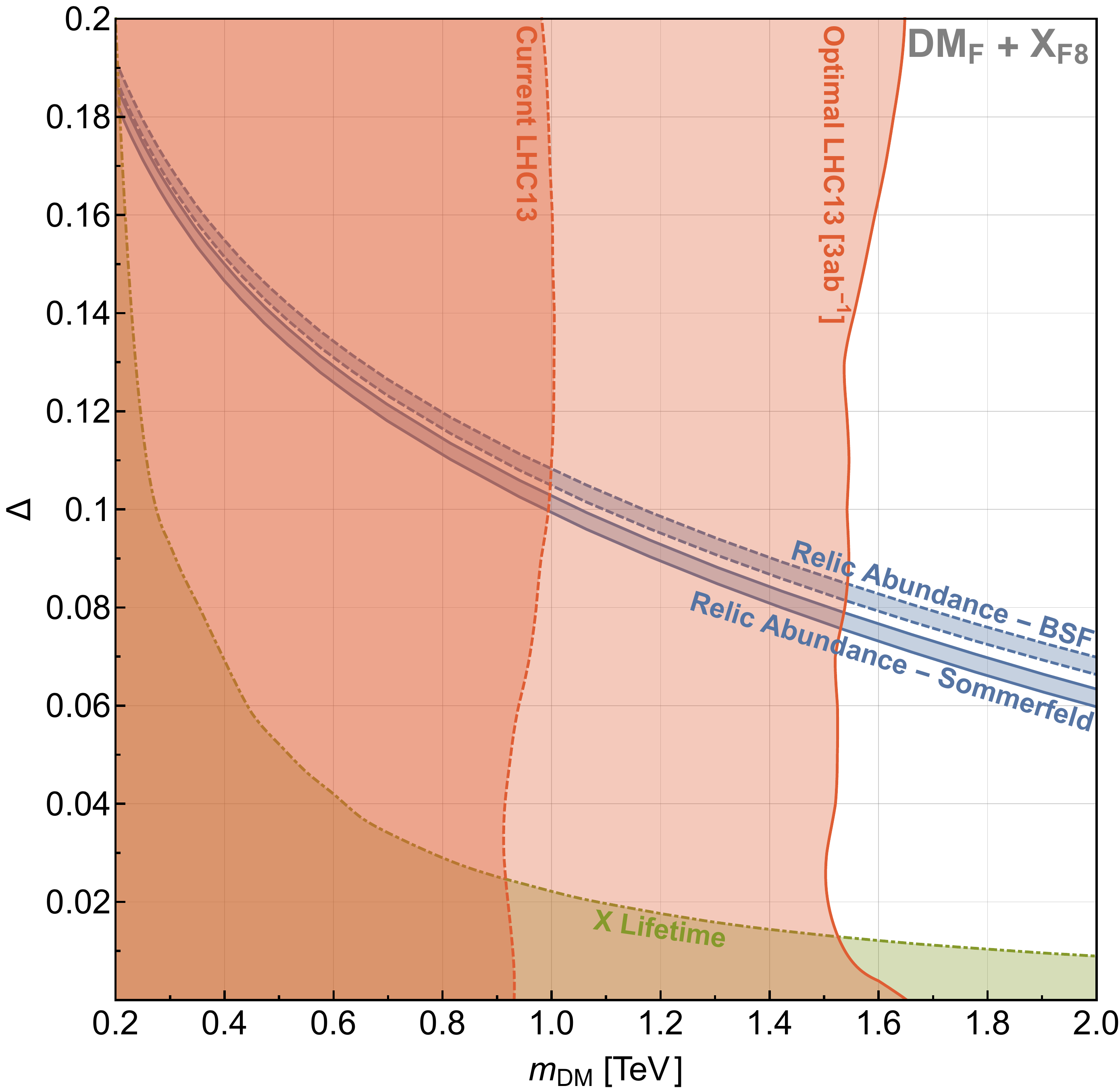}
	\includegraphics[width=0.48\textwidth]{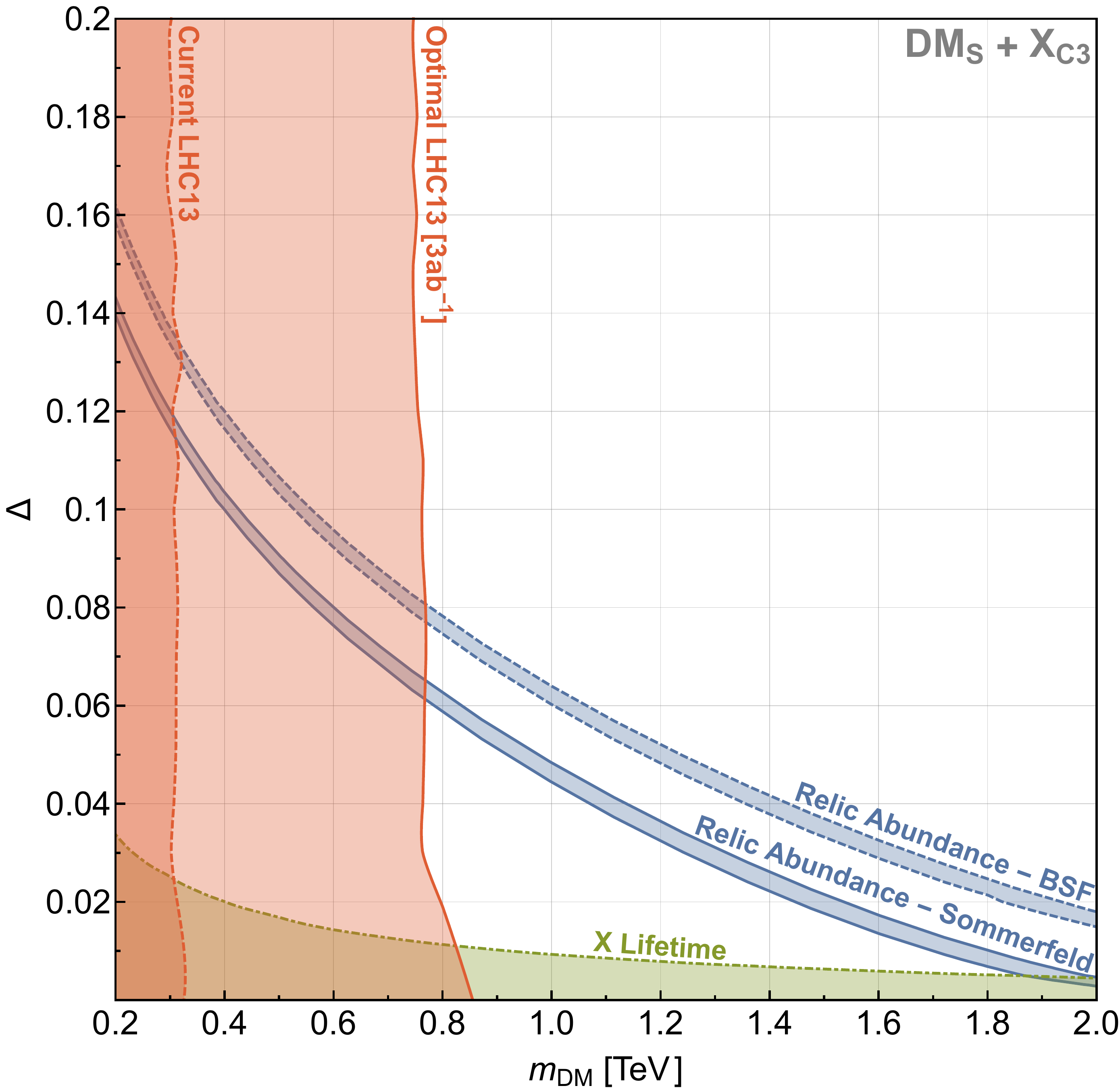}
	\includegraphics[width=0.48\textwidth]{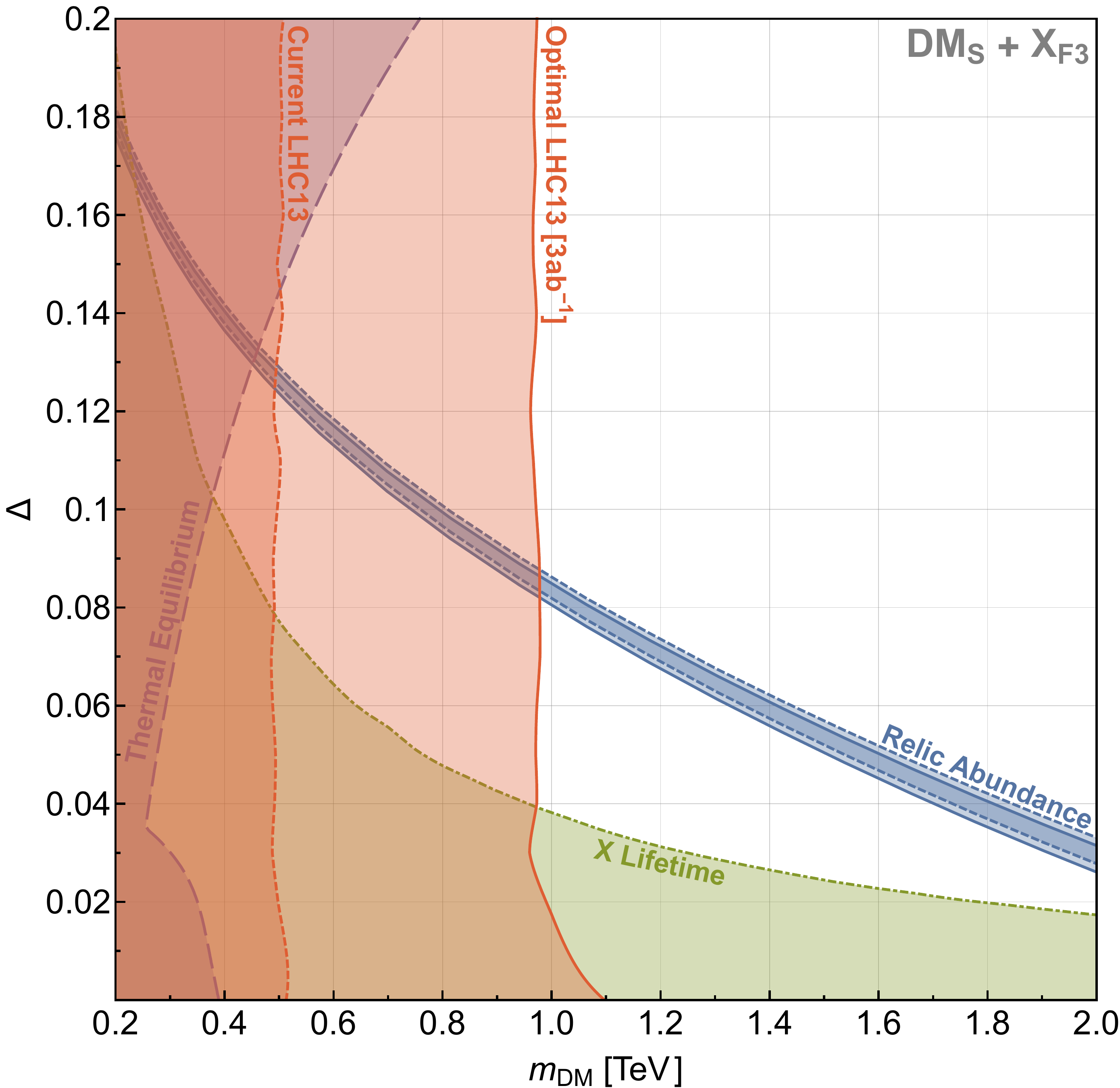}
	\includegraphics[width=0.48\textwidth]{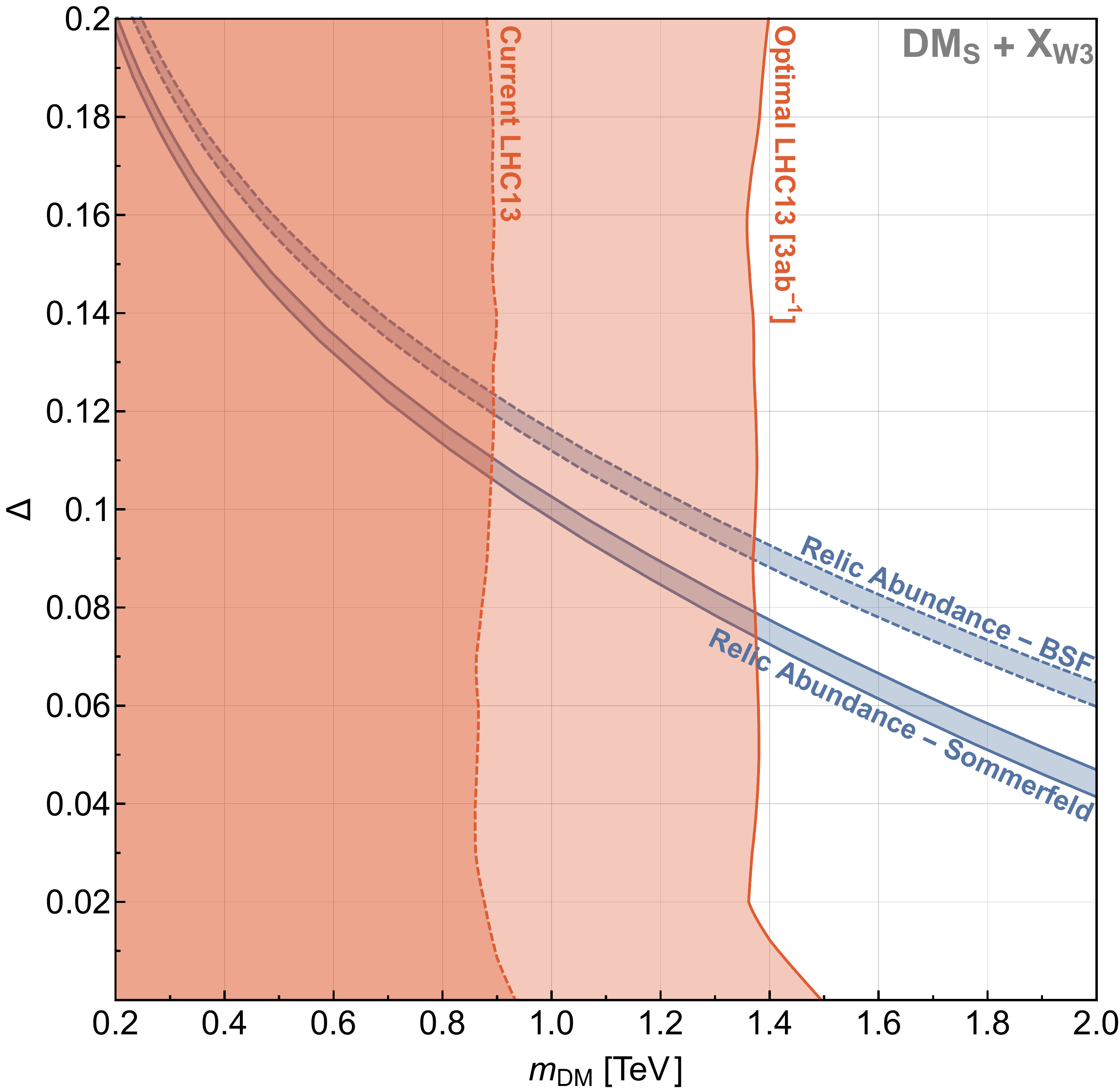}
	\caption{Constraints on our simplified models in the $(m_\mathrm{DM}, \Delta)$ plane. The blue regions give the correct relic abundance, including Sommerfeld corrections (solid) and bound states (dashed). In green the lifetime constraints on X are given assuming $3 \; \mathrm{ab}^{-1}$ at LHC13  and $\Lambda = 10$ TeV (see the main text for details). In red we show the current (dashed) and projected (solid) constraints from direct searches. The wide dashed line shows the thermal equilibrium bound for $\mathrm{DM}_\mathrm{S} + \mathrm{X}_\mathrm{F3}$.}
	\label{fig:combinations}
\end{figure}

\afterpage{\clearpage}

We show the LHC constraints on the $\Delta$ versus $m_\mathrm{DM}$ plane, in figure~\ref{fig:lhc:exclusion:bands} for the different models. For each $(m_\mathrm{DM}, \Delta)$ parameter point, we consider the signal region giving the highest significance. The exclusion bands range from the current bounds (derived using $S^{\rm 95}_{\rm obs}$, dashed boundary) up to an optimal end-of-lifetime LHC scenario for which the systematics are neglected and the total integrated luminosity is of $3000$ fb$^{-1}$ ($\beta=0$, solid boundary). We have explicitly verified that, keeping the current systematic errors for this increased luminosity only leads to a marginal gain with respect to the current results, and hence we do not show the corresponding bounds. For clarity reasons, we split the models into two sets. We present the fermionic $\mathrm{X}$ ($\mathrm{X}_\mathrm{F3}$, $\mathrm{X}_\mathrm{F6}$, $\mathrm{X}_\mathrm{F8}$) cases in the left panel, and all the models where $\mathrm{X}$ is a color triplet ($\mathrm{X}_\mathrm{F3}$, $\mathrm{X}_\mathrm{C3}$, $\mathrm{X}_\mathrm{W3}$) in the right panel. Note that the bounds on X are expected to be insensitive to the spin of the dark matter particle, and we have indeed verified that the $\mathrm{DM}_\mathrm{S} + \mathrm{X}_\mathrm{F3}$ and $\mathrm{DM}_\mathrm{F} + \mathrm{X}_\mathrm{F3}$ models yield the same exclusion curves.

From figure~\ref{fig:lhc:exclusion:bands} we observe that the bounds from the current ATLAS searches range from 300 GeV for the $\mathrm{X}_\mathrm{C3}$  model (that is for X being a complex scalar, color triplet) up to about 900 GeV for X being a fermion and either a color sextet or an octet. These values move up to about $750$ and $1500$ GeV respectively for the optimal high luminosity LHC scenario. Note that while the dashed boundaries are mostly vertical, the solid boundaries are vertical down to about a value of $\Delta \sim 1-2 \%$ beyond which the exclusion bound on $m_\mathrm{DM}$ increases as $\Delta$ becomes smaller. This small $\Delta$ region corresponds to a regime where the decay products of X escape detection most of the time, causing the monojet search to perform better than the multi-jet search due to lower statistical uncertainties. For larger values of the DM~--~X~splitting, the decay products of X become harder and the multi-jet search becomes more sensitive to the signal. Since this search allows for a large number of jets it is only weakly sensitive to $\Delta$, which accounts for the vertical exclusion bounds. 

Finally, in figure~\ref{fig:combinations}, we combine the information from the LHC exclusions with the parameter space regions fulfilling the relic density requirement as well as the lifetime and thermal equilibrium constraints on X for a luminosity of $3000~\mathrm{fb}^{-1}$. We first note that the current LHC searches set an upper bound on the required $\Delta$ ranging from $12$\% for the $\mathrm{DM}_\mathrm{S} + \mathrm{X}_\mathrm{C3}$ model down to $7.5$\% for the $\mathrm{DM}_\mathrm{F} + \mathrm{X}_\mathrm{F3}$ model. For $3000~\mathrm{fb}^{-1}$ luminosity and no systematics, these values shrink down to about $8 - 9$\% for most models, except for the $\mathrm{DM}_\mathrm{F} + \mathrm{X}_\mathrm{F3}$ model where the allowed value goes down to $4$\%. In addition, the lifetime constraints exclude the $\Delta < 1 - 2 \%$ region for the dark matter masses of interest here, thus generating a "wedge" in the parameter space, that the LHC would not be able to cover. The thermal equilibrium bound is relevant only for the $\mathrm{DM}_\mathrm{S} + \mathrm{X}_{\mathrm{F}3}$ model and excludes a large portion of the parameter space currently tested at the LHC. However, the LHC is already superseding this bound in the region consistent with relic density requirements. These results highlight the relevance of a proper determination of the relic density in order to correctly assess the status of the allowed parameter space.

Our study provides an important ingredient for future LHC analyses, setting the target parameter space to ${\cal O} (1-2)$~TeV dark matter masses with $\mathcal{O}(1-10) \%$ relative splittings with their colored coannihilation partners. This result can be used in conjunction with the simplified models we presented to design a tailored jets plus MET search at a hadron collider.

In view of the existing plans to construct a $100$~TeV collider, these simplified models for coannihilating dark matter once more stress the importance of continuing an experimental high-energy program to search for new physics. Such a collider could close the "wedge" shown in figure~\ref{fig:combinations} not only for the models presented here, but also for all 72 models with different spins and color charges. 

This "wedge" could be closed from the left side by direct searches, with a projected reach of about $3$ and $6$~TeV for compressed stops~\cite{Low:2014cba,Golling:2016gvc} and compressed gluinos~\cite{Low:2014cba} respectively. The small $\Delta$ region, namely the lower side of the wedge, will be increasingly constrained by the lifetime requirements on X. The increased reach is due to the higher X pair-production cross section, and by the fact that, in the absence of any signals of New Physics, the scale $\Lambda$ should be adjusted accordingly. Models where X is a fermion or a vector and a color sextet or octet can still satisfy the relic abundance constraints with dark matter masses slightly over $10$~TeV. Hence it is foreseeable that a ultra-heavy ($m \gtrsim 10$~TeV) and ultra-compressed ($\Delta \lesssim 1 \%$) region would be difficult to probe even with a 100 TeV collider. Such a scenario calls for dedicated strategies, for instance the use of specific jet reconstruction techniques for ${\cal O} (10\; \mathrm{GeV})$ jets, a modified detector with a pixel layer or a tracker closer to the beampipe, that would greatly enhance the naive expectations of the multi-jet + MET and long-lived particle analyses, respectively. We defer the study of the prospects of a higher energy collider and these difficult regions for future work.

%% file: sections/conclusions_v8.tex
\section{Conclusions}
\label{sec:conclusions}
In this work we studied the LHC exclusion reach for models where the dark sector includes not only the dark matter but also an additional colored field close in mass, generically dubbed X. Such models lead to coannihilation between the DM and its colored partner, the relic density being determined mainly by the processes $\mathrm{X} \, \overline{\mathrm{X}} \to q \, \bar{q}$, $g \, g$. The collider phenomenology of these scenarios is dominated by the pair production of X, followed by its decay into DM and additional jets. This decay proceeds via a higher-dimensional operator, which depends on the choices of quantum numbers of DM and X. We discuss the constraints associated to this operator, namely the possibility of X being long-lived, the perturbative unitarity bound for vector X as well as the thermal equilibrium requirement.

In this study, we reviewed the subtleties associated with a correct inclusion of the Sommerfeld effect and the contribution from bound states in the dark sector, following the treatment of reference~\cite{ElHedri:2016onc}. These two effects can lead to dramatic corrections to the thermal relic abundance and thus are necessary in order to test the thermal dark matter hypothesis.

We considered two different subsets of representative models. In the first one, we take DM and X to be Dirac fermions and study the effect of the color representation of X, which can be either a $\irrep{3}$, a $\irrep{6}$ or an $\irrep{8}$ of $SU(3)$. In the second one, we set X to be a color triplet and study the effects of its spin by taking it to be either a complex scalar, a complex vector or a Dirac fermion, with DM always being a real scalar. For these different models, we computed the allowed relic density regions in the $\Delta$ versus $m_\mathrm{DM}$ plane, where $\Delta$ is the fractional mass splitting between DM and X.

We also investigated the LHC constraints on the aforementioned models obtained by recasting the ATLAS searches for mono and multi-jet plus MET. We do not only derive the current limits on the $(m_\mathrm{DM}, \Delta)$ parameter space but also compute the projected bounds expected at HL-LHC for a luminosity of $3000~\mathrm{fb}^{-1}$. We perform an estimate of the current systematic uncertainties, and show that extrapolating the current studies to this higher luminosity would only marginally increase the reach in mass compared to the current results, unless the systematic errors can be significantly reduced. In addition to the current LHC bounds, we therefore present the limits associated to an ``optimal'' HL-LHC scenario where the systematic errors are set to zero. This optimal configuration can lead to a factor of two improvement of the current limits on the dark matter mass. The allowed mass splittings $\Delta$ can typically be reduced by the same amount although the associated bounds are highly model-dependent since they require interfacing the LHC results with the relic density constraints. 

Thermal dark matter models provide a compelling and elegant explanation for the current observed dark matter relic density but are also being increasingly constrained by the current collider, direct and indirect detection experiment. In this context, scenarios where the dark matter coannihilates with a strong partner are becoming one of the rare viable options for multi-TeV dark matter. In this work, we showed that the LHC would be able to probe most of the regions with a ``natural'' mass splitting between the dark matter and its partner. 

The regions of the parameter space with $\Delta \lesssim 10$\% are typically associated with multi-TeV dark matter masses and can only be probed by a collider with higher center-of-mass energy than the LHC. The next generation of particle accelerators could therefore be instrumental in probing one of the last remaining thermal dark matter scenarios. Existing studies suggest that a prospective $100$ TeV collider would be able to considerably narrow down the parameter space of models of strongly coannihilating dark matter, potentially leaving a window in the ultra-compressed ($\Delta \lesssim 1$\%) and ultra-heavy ($m_{\mathrm{DM}} \gtrsim 10$ TeV) region. Such a window could be accessed with a refinement of the analysis techniques and improvement in the detector design. Hence colored dark sectors with thermal dark matter provide an ideal physics case for the development of future particle colliders.

%% file: sections/acknowledgements_v5.tex
\acknowledgments
We would like to thank Seng Pei Liew, Andreas Papaefstathiou and Felix Yu for valuable discussions. This research is supported by the Cluster of Excellence Precision Physics, Fundamental Interactions and Structure of Matter (PRISMA-EXC 1098), by the ERC Advanced Grant EFT4LHC of the European Research Council, and by the Mainz Institute for Theoretical Physics.

%% file: sections/thermal_eq_v7.tex
\section{Thermal equilibrium}
\label{sec:thermal:eq:calc}
In this appendix we derive the DM~$\leftrightarrow$~X conversion rates for $\mathrm{DM}_\mathrm{S} + \mathrm{X}_\mathrm{F3}$ model, whose DM\,X\,SM$_1$\,SM$_2$ operator from equation~\eqref{eq:lagrangians:eft} is suppressed by both two powers of $\Lambda$ and a loop factor. We have verified that this model is the only one in our study that is associated with non-trivial constraints from thermal equilibrium, due to the additional loop suppression. Following~\cite{Ellis:2015vaa} as in section~\ref{sec:thermal:eq}, we require that the inequality~\eqref{eq:thermal:eq}, where the thermally averaged rate is given by equation~\eqref{eq:thermal:averaged:rate}, is satisfied. Here, we provide the scattering cross sections that should be injected in~\eqref{eq:thermal:eq} for the $\mathrm{DM} \, g \rightarrow \mathrm{X} \, \bar{d}$, $\mathrm{DM} \, d \rightarrow \mathrm{X} \, g$, $\mathrm{X} \, g \rightarrow \mathrm{DM} \, d$ and $\mathrm{X} \, \bar{d} \rightarrow \mathrm{DM} \, g$ processes. The thermal equilibrium condition for processes involving $\overline{\mathrm{X}}$ will lead to exactly the same constraints.

To obtain $\sigma(s)$ we work in the center-of-mass frame where the total scattering cross section is calculated using
\begin{equation} \label{eq:sigma:thermal}
	\sigma(s) = \frac{p_f}{16 p_i s (2 \pi)^2} \int | \overline{\mathcal{M}} |^2 \mathrm{d} \Omega ,
\end{equation}
where $p_i = |\vec{p}_i|$ and $p_f = |\vec{p}_f|$ are the momenta of the initial and final state particles respectively. In here $| \overline{\mathcal{M}} |^2$ is the spin and color averaged squared matrix element for the different processes. For the $\mathrm{DM}_\mathrm{S} + \mathrm{X}_\mathrm{F3}$ the color factor is $4$ and together with averaging over the spin and color degrees of freedom of the initial states we obtain a prefactor for each of the processes
\begin{equation}
	\mathcal{C}_{\mathrm{DM} \, g \rightarrow \mathrm{X} \, \bar{d}} = \frac{1}{4} \qquad \mathcal{C}_{\mathrm{DM} \, d \rightarrow \mathrm{X} \, g} = \frac{2}{3} \qquad \mathcal{C}_{\mathrm{X} \, g \rightarrow \mathrm{DM} \, d} = \frac{1}{24} \qquad \mathcal{C}_{\mathrm{X} \, \bar{d} \rightarrow \mathrm{DM} \, g} = \frac{1}{9}.
\end{equation}
With these prefactors the cross sections for the different two-to-two processes responsible for attaining thermal equilibrium are
\begin{equation}
	\begin{aligned}
		\sigma_{\mathrm{DM} \, g \rightarrow \mathrm{X} \, \bar{d}}(s) & =  \frac{p_f}{32\pi \, p_i \, s}\,\frac{\mathcal{C}_{\mathrm{DM} \, g \rightarrow \mathrm{X} \, \bar{d}}}{(16\pi^2\Lambda^2)^2} \,\frac{8 (s - m_\mathrm{DM}^2)^2(s-m_\mathrm{X}^2)(2m_\mathrm{X}^2 + s)}{3s}\\
		\sigma_{\mathrm{DM} \, d \rightarrow \mathrm{X} \, g}(s) & =  \frac{p_f}{32\pi \, p_i \, s}\,\frac{\mathcal{C}_{\mathrm{DM} \, d \rightarrow \mathrm{X} \, g}}{(16\pi^2\Lambda^2)^2} \,\frac{8 (s - m_\mathrm{X}^2)^2(s-m_\mathrm{DM}^2)}{s}\\
		\sigma_{\mathrm{X} \, g \rightarrow \mathrm{DM} \, d}(s) & =  \frac{p_f}{32\pi \, p_i \, s}\,\frac{\mathcal{C}_{\mathrm{X} \, g \rightarrow \mathrm{DM} \, d}}{(16\pi^2\Lambda^2)^2} \,\frac{8 (s - m_\mathrm{X}^2)^2(s-m_\mathrm{DM}^2)}{s}\\
		\sigma_{\mathrm{X} \, \bar{d} \rightarrow \mathrm{DM} \, g}(s) & =  \frac{p_f}{32\pi \, p_i \, s}\,\frac{\mathcal{C}_{\mathrm{X} \, \bar{d} \rightarrow \mathrm{DM} \, g}}{(16\pi^2\Lambda^2)^2} \,\frac{8 (s - m_\mathrm{DM}^2)^2(s-m_\mathrm{X}^2)(2m_\mathrm{X}^2 + s)}{3s}.
	\end{aligned}
\end{equation}
In here the initial and final state momenta are given by
\begin{equation}
    p_i = \frac{s - m_i^2}{2 \sqrt{s}} \qquad \quad p_f = \frac{s - m_f^2}{2 \sqrt{s}} ,
\end{equation}
where $m_{i,f}$ are the masses of the dark sector particles (DM or X) in the initial and final state, respectively. The constraints for the other models can be obtained in a similar fashion by calculating the respective squared matrix elements.

%% file: sections/bound_states_v7.tex
\section{Bound state dynamics}
\label{sec:bound:states}
This section outlines how to compute the contributions from bound state formation and decay to the dark matter effective annihilation cross section. In particular, we show how to compute the bound state formation and dissociation cross sections as well as their decay rates in the non-relativistic limit. Here, we follow the procedure introduced in~\cite{Liew:2016hqo} and focus on color-singlet $\ell = 0$ bound states.\footnote{For a study considering the effects of $p$-wave bound states, thermal corrections, the non-Abelian structure of QCD and bound states in the octet representation, see reference~\cite{Mitridate:2017izz}.} We extend the results in~\cite{Liew:2016hqo} for scalar and fermion color triplets and octets to vectors and color sextets as well. Due to the symmetry requirements on their wave function these bound states need to have even spins. Therefore, in what follows, we restrict ourselves to bound states of spin $0$ for scalar and fermion X and of spin $0$ and $2$ for vector X.

A given X$\overline{\mathrm{X}}$ bound state can dissociate into X and $\overline{\mathrm{X}}$ by absorbing a gluon. As shown in~\cite{Liew:2016hqo}, the corresponding dissociation cross sections are independent of the spin of X and the bound state. They factorize into a color-independent term times symmetry and color factors in the following way
\begin{equation}
	\begin{aligned}
		\sigma_\mathrm{dis}^{S3, F3, V3} & = \frac{1}{8} \times \frac{4}{3} \times \sigma^0_{\mathrm{dis}, r}\\
		\sigma_\mathrm{dis}^{S6, F6, V6} & = \frac{1}{8} \times \frac{10}{3} \times \sigma^0_{\mathrm{dis}, a}\\
		\sigma_\mathrm{dis}^{S8, F8, V8} & = \frac{1}{8} \times 3 \times \left( 2 \; \textrm{for identical particles} \right) \times \sigma^0_{\mathrm{dis}, a}.
	\end{aligned}
\end{equation}
Here, the subscripts $a$ and $r$ indicate whether the QCD potential between the two final state particles given in equation~\eqref{eq:potential:qcd} is attractive or repulsive. The $\sigma^0_{\mathrm{dis}, a}$ and $\sigma^0_{\mathrm{dis}, r}$ cross sections can be written as
\begin{equation}
	\begin{aligned}
		\sigma^0_\mathrm{dis,a} &= \frac{2^9\pi^2}{3} \alpha_s a^2 \left(\frac{E_B}{\omega}\right)^4 \frac{1 + \nu^2}{1 + (\kappa\nu)^2} \frac{e^{-4\nu\mathrm{arccot}(\kappa\nu)}}{1 - e^{-2\pi\nu}}\kappa^{-1}\\
		\sigma^0_\mathrm{dis,r} &= \frac{2^9\pi^2}{3} \alpha_s a^2 \left(\frac{E_B}{\omega}\right)^4 \frac{1 + \nu^2}{1 + (\kappa\nu)^2} \frac{e^{-4\nu\mathrm{arccot}(\kappa\nu) - 2\pi\nu}}{1 - e^{-2\pi\nu}}\kappa^{-1} ,
	\end{aligned}
\end{equation}
where $v_{rel}$ is the relative velocity between the two outgoing particles, $\omega$ is the energy of the incoming gluon, and we define 
\begin{equation}
	\nu = \frac{|\zeta'|}{v_{rel}} \qquad \kappa = \frac{\zeta}{|\zeta'|} \qquad a = (\zeta \mu)^{-1} \qquad E_B = \frac{\zeta^2\mu}{2} . 
\end{equation}
$E_B$ and $a$ are the binding energy and the Bohr radius respectively, whereas $\mu = m_X/2$ is the reduced mass of the two-particle system. The modified couplings $\zeta$ and $\zeta'$ respectively associated to the bound state and the final two-particle state are given in equation~\eqref{eq:alpha}. Since we always consider color-singlet bound states, the X$\overline{\mathrm{X}}$ pair in the final state will always be a color octet. We can therefore write
\begin{equation}
	\zeta = C_\mathrm{X}\alpha_s\qquad \zeta' = \left(C_\mathrm{X} - \frac{3}{2}\right) \alpha_s ,
\end{equation}
where $C_\mathrm{X}$ is the quadratic Casimir index of the color representation of X and is equal to $\frac{4}{3}, \frac{10}{3}$ and $3$ for triplet, sextet and octet X respectively.

The formation cross sections can be obtained from the dissociation cross section by
\begin{equation}
	\sigma_\mathrm{bsf} = \left( 2 \; \textrm{for identical particles} \right) \times \frac{g_\mathrm{BS} g_g}{g_{X}^2} \left( \frac{\omega}{\mu v_{rel}} \right)^2 \sigma_\mathrm{dis} ,
\end{equation}
where $g_\mathrm{BS}$ is the number of degrees of freedom of the bound state, $g_g = 16$ is the number of degrees of freedom of the gluon, and $g_X = d_R \times (2 s + 1)$ is the number of degrees of freedom of X. For each model, we also have to take into account the fact that not all of the initial state degrees of freedom will contribute to the formation of a bound state with a given spin. For spin $0$ bound states, the bound state formation cross section should therefore be multiplied by $1/4$ and $1/9$ for fermions and vectors respectively while for spin 2 bound state, the cross section for vectors should be multiplied by $5/9$. The bound state formation cross sections for scalars are left unchanged.

Bound states can decay via either the individual decay of their components or their annihilation. The second process is largely dominating and, for an $s$-wave bound state of spin $s$, leads to the following width~\cite{Kahawala:2011pc}
\begin{equation}
	\begin{aligned}
		\Gamma = & \frac{\zeta^3 m_X}{1024 \pi^3} \times \mathcal{C}_1 \times \sum_{m_{g_1}m_{g_2}} |\mathcal{M}(0;00;ss_z;m_{g_1} m_{g_2})|^2 \\
		& \times \left(\frac{1}{2} \; \textrm{for identical bound state constituents} \right) \\
		& \times \left(\frac{1}{2} \; \textrm{for identical final state particles} \right) ,
	\end{aligned}
\end{equation}
where $\mathcal{M}(0;00;s s_z;m_{g_1} m_{g_2})$ is the $l = 0$ component of the $\mathrm{X} \, \overline{\mathrm{X}} \rightarrow q \, \bar{q}, g \, g$ perturbative amplitude for $v\rightarrow 0$. $m_{g_1}$ and $m_{g_2}$ are the z-components of the final state spins. $\mathcal{C}_1$ is the fraction of the particle-antiparticle state that ends up in a color singlet. For our models, the widths of the bound states are then
\begin{equation}
	\begin{aligned}
		& \Gamma^{S3} = \frac{1}{3} \mu \alpha_s^2 \zeta^3 \quad && \Gamma^{F3} = \frac{2}{3} \mu \alpha_s^2 \zeta^3 \quad && \Gamma^{V3}_{s=0} = \mu \alpha_s^2 \zeta^3 \quad && \Gamma^{V3}_{s=2} = \frac{16}{3}\mu \alpha_s^2 \zeta^3 \\
		& \Gamma^{S6} = \frac{25}{6} \mu \alpha_s^2 \zeta^3 \quad && \Gamma^{F6} = \frac{25}{3} \mu \alpha_s^2 \zeta^3 \quad && \Gamma^{V6}_{s=0} = \frac{25}{2} \mu \alpha_s^2 \zeta^3 \quad && \Gamma^{V6}_{s=2} = \frac{200}{3} \mu \alpha_s^2 \zeta^3 \\
		& \Gamma^{S8} = \frac{9}{2} \mu \alpha_s^2 \zeta^3 \quad && \Gamma^{F8} = 9 \mu \alpha_s^2 \zeta^3 \quad && \Gamma^{V8}_{s=0} = \frac{27}{2} \mu \alpha_s^2 \zeta^3 \quad && \Gamma^{V8}_{s=2} = 72 \mu \alpha_s^2 \zeta^3 .
	    \label{eq:decay}
	\end{aligned}
\end{equation}
Here, we assumed that X and $\overline{\mathrm{X}}$ are not the same. For the octet, if X is its own antiparticle the rates have to be divided by two.

The averaged bound state dissociation rates can be expressed as a function of the dissociation cross section $\sigma_\mathrm{dis}$ in the following way
\begin{equation}
	\langle \Gamma \rangle_\mathrm{dis} = g_g \frac{4 \pi}{(2 \pi)^3} \int_0^\infty \mathrm{d} u \, \sigma_\mathrm{dis} \frac{E_B^3 \left(1 + \frac{u}{z} \right)^2}{z (e^{z + u} - 1)} ,
\end{equation}
where $u \approx \frac{1}{2} \mu v_{rel}^2 / T$ and $z = E_B / T$. The thermally-averaged bound state decay rates are proportional to the $\Gamma_\eta$ bound state decay rates given in equation~\eqref{eq:decay} for each model and are given by
\begin{equation}
	\langle \Gamma \rangle_\eta = \Gamma_\eta \frac{K_1 (m_\eta / T)}{K_2 (m_\eta / T)} ,
\end{equation}
where $m_\eta$ is the mass of the bound state and $K_1$, $K_2$ are modified Bessel functions of the second kind. These two quantities have been computed in the non-relativistic limit. Finally, we compute the thermally-averaged bound state formation cross section. Here, we use the \texttt{micrOMEGAs} code~\cite{Belanger:2014vza} and perform the following relativistic averaging
\begin{equation}
	\begin{aligned}
		\langle \sigma v \rangle_\mathrm{bsf}  & = \frac{T}{8 \pi^4 \bar{n} (T)^2} \int \mathrm{d} s \, \sqrt{s} K_1 \left( \frac{\sqrt{s}}{T} \right) g_\mathrm{X}^2 \left(s - \frac{m_\mathrm{X}^2}{4} \right) \left(1 + \frac{1}{e^{\omega / T} - 1} \right) \sigma_\mathrm{bsf} (s) \\
		\bar{n} (T) & = \frac{T}{2 \pi^2} \left[g_\mathrm{X}^2 m_\mathrm{X}^2 K_2 \left(\frac{m_\mathrm{X}}{T} \right) + g_\mathrm{DM}^2 m_\mathrm{DM}^2 K_2 \left( \frac{m_\mathrm{DM}}{T} \right) \right] ,
	\end{aligned}
\end{equation}
where $s$ is the center-of-mass energy of the collision. The factor $\frac{1}{e^{\omega / T} - 1}$ corresponds to the enhancement of the bound state formation rate from the stimulated emission due to the gluons in the thermal bath. The final thermally-averaged dark matter annihilation cross section including bound state effects can then be written as
\begin{equation}
	\langle \sigma v \rangle_\mathrm{ann} = \langle \sigma v \rangle_\mathrm{pert} + \langle \sigma v \rangle_\mathrm{bsf} \frac{\langle \Gamma \rangle_\eta}{\langle \Gamma \rangle_\eta + \langle \Gamma \rangle_\mathrm{dis}}.
\end{equation}
We updated the package~\cite{ElHedri:2016pac} to include these bound state formation effects, using a slightly modified version of \texttt{micrOMEGAs}. Note that since we perform a relativistic thermal averaging for $\sigma_\mathrm{bsf}$, our results are slightly different from the ones presented in \cite{Liew:2016hqo} that are derived under the non-relativistic approximation.